\RequirePackage{fix-cm}
\documentclass{svjour3}                     
\smartqed  
\usepackage{amsmath}
\usepackage{amssymb}
\usepackage{graphicx}
\usepackage{booktabs}
\usepackage{natbib}
\usepackage{color}
\usepackage{hyperref}
\usepackage{subcaption}
 \journalname{arXiv}
\begin{document}

\title{An analytical solution for horizontal velocity profiles in the hurricane boundary layer}

\titlerunning{Analytical solution for the HBL velocity profiles}  

\authorrunning{K.~R.~Sathia $\cdot$  M.~G.~Giometto} 
\author{K.~R.~Sathia\and M.~G.~Giometto$^\dagger$}

\institute{ 
K.~R.~Sathia \at Department of Civil Engineering and Engineering Mechanics, Columbia University, New York, NY, USA \\
		\email{krs2199@columbia.edu}
\and
$\dagger$ M.~G.~Giometto \at Department of Civil Engineering and Engineering Mechanics, Columbia University, New York, NY, USA \\
		\email{mg3929@columbia.edu}
}

\date{Received: DD Month YEAR / Accepted: DD Month YEAR}

\maketitle

\begin{abstract}
Theoretical analyses of the hurricane boundary layer have traditionally relied on slab models, which provide a limited description of wind profiles. 
Literature on height-resolving methods is typically based on linear analyses, which may fall short of capturing the full sensitivity of the solution to variations in input parameters.
This work proposes an approximate analytical solution of a nonlinear single-column model for horizontal winds based on a constant eddy viscosity assumption and valid outside the eyewall.
Building on literature that uses a linearized system of equations, we account for the nonlinearities through a series expansion approach. 
We find that a first order correction is sufficient for most practical cases. 
This solution helps provide a simplified understanding of the sensitivity of the radial and tangential wind profiles to input parameters such as the distance from the hurricane eye and the Coriolis force.


\end{abstract}

\section{Introduction}
The vertical structure of mean wind in hurricane boundary layers (HBLs) is of relevance for a wide range of engineering applications, including hurricane forecasting, damage modeling, and risk assessment for coastal infrastructure. 

Practical difficulties in obtaining wind speed and flux measurements in the HBL have traditionally limited the amount of available observational data. 
Advances over the past two decades in observational capabilities \citep{cione2020eye} have provided significant insight into HBL processes and informed the development of empirical wind profiles such as those presented in \citet{vickery2009hurricane}, which have proved useful for operational use. 
Additionally, recent developments in turbulence-resolving HBL simulations \citep[e.g.,][]{bryan2017eddy, bryan2017simple, worsnop2017spectral, momen2021scrambling, chen2021framework, ma2021large}, have opened the doors for a more mechanistic understanding of HBL dynamics. 

Simplified theoretical analyses are also being explored that build on these advancements and enhance our understanding of the wind variation seen in numerical and observational studies. Theoretical analyses of the HBL are typically performed using slab models \citep{kepert2010slab, smith2023tropical} which average out or prescribe the vertical dynamical structure. 
Height-resolving analyses, with a focus on the velocity profiles, gained traction towards the last quarter of the 20th century \citep{rosenthal1962theoretical, smith1968surface, eliassen1977ekman, meng1995analytical}. 
A summary of these contributions is presented in \citet{chang2024comprehensive}.

An influential paper by \citet{kepert2001dynamics} incorporated much of the work in these previous studies and established the system of linear equations used in height-resolving models.
These equations are the radial and tangential momentum balance that use the gradient wind balance to model the radial pressure gradient, and neglect second and higher order products of velocities. 
In the limit of large radial distance from the storm center, the solution to these equations converges to the classic Ekman layer solution.
More recent publications built on this work to include vertically varying eddy viscosity \citep{vickery2009hurricane}, non-axisymmetric gradient wind \citep{snaiki2017linear} and the vertical advection in the eyewall \citep{yang2021height}. 
These works have provided insight into the vertical structure of the mean wind in the HBL but are based on a linearized formulation of the problem.
While there is some literature that discusses the drawbacks of neglecting nonlinearities \citep{kepert2001nonlinear, foster2009boundary, vogl2009limitations}, these analyses have relied on a numerical approach to solve the differential equations. 
There remains a need for an analytical formula for the variation of mean speed that is backed by nonlinear theory, matches realistic hurricane wind profiles, and is sufficiently simple for use in engineering practice. 
Such a solution would also help elucidate dependencies between flow features and make these easier to interpret.

This work proposes an analytical solution of the nonlinear HBL equations valid outside the eyewall, using a single column model with constant eddy-viscosity.
Building on \citet{kepert2001dynamics}, which analyses the linearised version of the same model, we use a series expansion to account for the nonlinearities.
We first discuss the governing equations, boundary conditions and the physical validity of the solutions to these equations in Sect. ~\ref{section:govEqns} and ~\ref{section:physValidity}. 
This is followed by the description of the series solution, of the nonlinear effects and a comparison against the numerical solution in Sect.~\ref{section:analyticalSoln}. The analytical solution allows for a simple analysis of sensitivity to the various parameters. This is explored in Sect.~\ref{section:sensAnalysis}. 
A simpler alternative solution for the velocity magnitude is proposed in Sect.~\ref{section:velMag}, which is informed by the analysis in Sect.~\ref{section:analyticalSoln} and ~\ref{section:sensAnalysis}.
Conclusions are drawn in Sect.~\ref{section:summary}.

\vfill
 
\section{Governing Equations and Non-Dimensionalization}\label{section:govEqns}
The governing equations for the mean velocity profiles in the stationary and axisymmetric HBL far from the eyewall are
\begin{eqnarray} \label{eq:toBeSolvedDimensional1}
\frac{d}{d z}\left(K\frac{d \overline{u}_r}{d z} \right) &=&
\left(\frac{V_g^2}{R} + fV_g\right) 
- \frac{\overline{u}^2_\theta}{R} 
- \frac{\overline{u}_r^2}{R}
- f\overline{u}_\theta \ ,
\\
\label{eq:toBeSolvedDimensional2}
\frac{d}{dz}\left(K\frac{d \overline{u}_\theta}{d z} \right)
&=&
\frac{\overline{u}_r\overline{u}_\theta}{R}
+ f\overline{u}_r 
- n\overline{u}_r\frac{V_g}{R} \ ,
\end{eqnarray} 
where $\overline{u}_r$ is the radial velocity, $\overline{u}_\theta$ is the tangential velocity, $V_g$ is the gradient wind speed, $f$ is the Coriolis frequency, $R$ is the radial distance from the axis of rotation, $n$ is the non-dimensional radial derivative of gradient wind and $K$ is the sum of the molecular and turbulent viscosities, which we treat in this analysis to be constant. A detailed derivation is provided in Appendix \ref{section:appendixGovEqns}. We have used the modelling choices $ \partial \overline{u}_r/\partial r = -\overline{u}_r/R$ and $
    \partial \overline{u}_\theta/\partial r = -nV_g/R$ which restrict the equations to the region outside the eyewall.
    
We start by considering a constant gradient wind $V_g = G$.
A linearly varying gradient wind is considered in Appendix \ref{section:appendixlinVarGradWind}. 
For a constant gradient wind, \eqref{eq:toBeSolvedDimensional1} and \eqref{eq:toBeSolvedDimensional2} is solved with the boundary conditions
\begin{equation}\label{eq:dimensionalBCs}
    \overline{u}_r(0) = \overline{u}_r(\infty) =  \overline{u}_\theta(0) = 0 \ ,  \qquad
    \overline{u}_\theta(\infty) = G \ .
\end{equation}

Consider the coordinate transformation 
$\hat{u} = \overline{u}_r$ and $\hat{v} = \overline{u}_\theta - G$. 
Next, introduce non-dimensional height $\xi = z/H$ and non-dimensional velocity variables $u = \hat{u}/G$ and $v = \hat{v}/G$. A depth scale similar to the Ekman layer depth $H = \sqrt{2K/I}$
can be constructed to normalize $z$, where,
\begin{equation} \label{eq:inertialStability}
    I = \sqrt{\left(f + \frac{2G}{R}  \right)\left(f + (1-n)\frac{G}{R}  \right)}
\end{equation}
is the inertial stability, as defined in \cite{kepert2001dynamics}. Substituting the transformations into \eqref{eq:toBeSolvedDimensional1} and \eqref{eq:toBeSolvedDimensional2}, one obtains
\begin{eqnarray} \label{eq:nonDimmedSimple1}
\frac{d^2 u}{d \xi^2} &=&
\frac{H^2G}{K R}\left[
- \left(\frac{1}{Ro} + 2\right)v
- (u^2 + v^2)
\right] \ ,
\\
\label{eq:nonDimmedSimple2}
\frac{d^2 v}{d \xi^2} &=&
\frac{H^2G}{K R}\left[
\left(\frac{1}{Ro} + (1-n)\right)u
+
uv
\right] \ ,
\end{eqnarray}
where $Ro = G/(fR)$ is a Rossby number based on the gradient wind speed $G$ and the radial distance $R$ at which the velocity profiles are to be determined. Substituting for $H$, the equations become
\begin{eqnarray}\label{eq:nonDimmedEqns1}
\frac{d^2 u}{d \xi^2} &=&
- \tilde{\alpha} v
- \tilde{\gamma}(u^2 + v^2) \ ,
\\
\label{eq:nonDimmedEqns2}
\frac{d^2 v}{d \xi^2} &=&
\tilde{\beta} u+
\tilde{\gamma} uv \ ,
\end{eqnarray}
where
\begin{equation} \label{eq:nonDimmedParams}
    \begin{gathered}
\tilde{\alpha} = 2\sqrt{\frac{\left(\frac{1}{Ro} + 2 \right)}{\left(\frac{1}{Ro} + (1-n)\right)}} \ , \qquad \qquad \qquad
\tilde{\beta} = 2\sqrt{\frac{\left(\frac{1}{Ro} + (1-n)\right)}{\left(\frac{1}{Ro} + 2 \right)}} \ , \\
\tilde{\gamma} = \frac{2}{\sqrt{\left(\frac{1}{Ro} + 2\right)\left(\frac{1}{Ro} + (1-n)\right)}} \ .
    \end{gathered}
\end{equation}
Note that $\{\tilde{\alpha},\tilde{\beta},\tilde{\gamma}\} = 
2H^2\{\alpha, \beta, \gamma\}$ where $\alpha$, $\beta$ and $\gamma$ are defined in Equation 8 of \cite{kepert2001dynamics}.
Normalized boundary conditions for \eqref{eq:nonDimmedEqns1} and \eqref{eq:nonDimmedEqns2} are
\begin{equation} \label{eq:nonDimmedBCs}
\begin{gathered}
     u(0) = u(\infty) = v(\infty) = 0\ , \qquad v(0) = -1\  .
\end{gathered}
\end{equation}

We emphasize here that the non-dimensional problem depends only on two parameters, namely $n$ and the non-dimensional Coriolis frequency $1/Ro$.
\section{Physical validity of governing equation}\label{section:physValidity}
Hurricanes occur in the Northern Hemisphere roughly in the latitudes $8^\circ$N to $40^\circ$N and with gradient wind speeds from about $20$ to $80$ m/s.
``Outside the eyewall" could be considered to be $R\in [20,200]$ km. 
These values correspond to $Ro \in [1, 200]$. 
The only other parameter in the non-dimensional equations is $n$. 
From Table 2 of \cite{mallen2005reexamining}, $n \in [0.04,0.67]$.

Consider three cases of $n$, as tabulated in Table~\ref{tab:n_cases}.
Equations \eqref{eq:nonDimmedEqns1} and \eqref{eq:nonDimmedEqns2} with boundary conditions \eqref{eq:nonDimmedBCs} are solved over a finite but sufficiently large domain numerically for cases 1 and 3 and are shown in Fig.~\ref{fig:n_cases_anlnumSoln}.

\begingroup
    \setlength{\tabcolsep}{12pt}
\renewcommand{\arraystretch}{1.5}
\begin{table}
    \centering
    \begin{tabular}{ccccc}
    \hline
          &$\mathbf{n}$ & $\mathbf{\tilde{\alpha}}$ & $\mathbf{\tilde{\beta}}$ & $\mathbf{\tilde{\gamma}}$ \\
         \hline 
        \textbf{Case 1} & 0.30 & 3.37 &
1.19 & 1.67 \\
        \textbf{Case 2} & 0.45 & 3.79 &
1.06 & 1.89 \\
        \textbf{Case 3} & 0.60 & 4.43 &
0.90 & 2.20 \\
        \hline
    \end{tabular}
    \caption{Three cases analysing the effect of $n$ (with $Ro = 100$).}
    \label{tab:n_cases}
\end{table}
\endgroup

\begin{figure}[htpb]
    \centering
    \begin{subfigure}[b]{0.49\textwidth} 
        \centering
        \includegraphics[width=\textwidth]{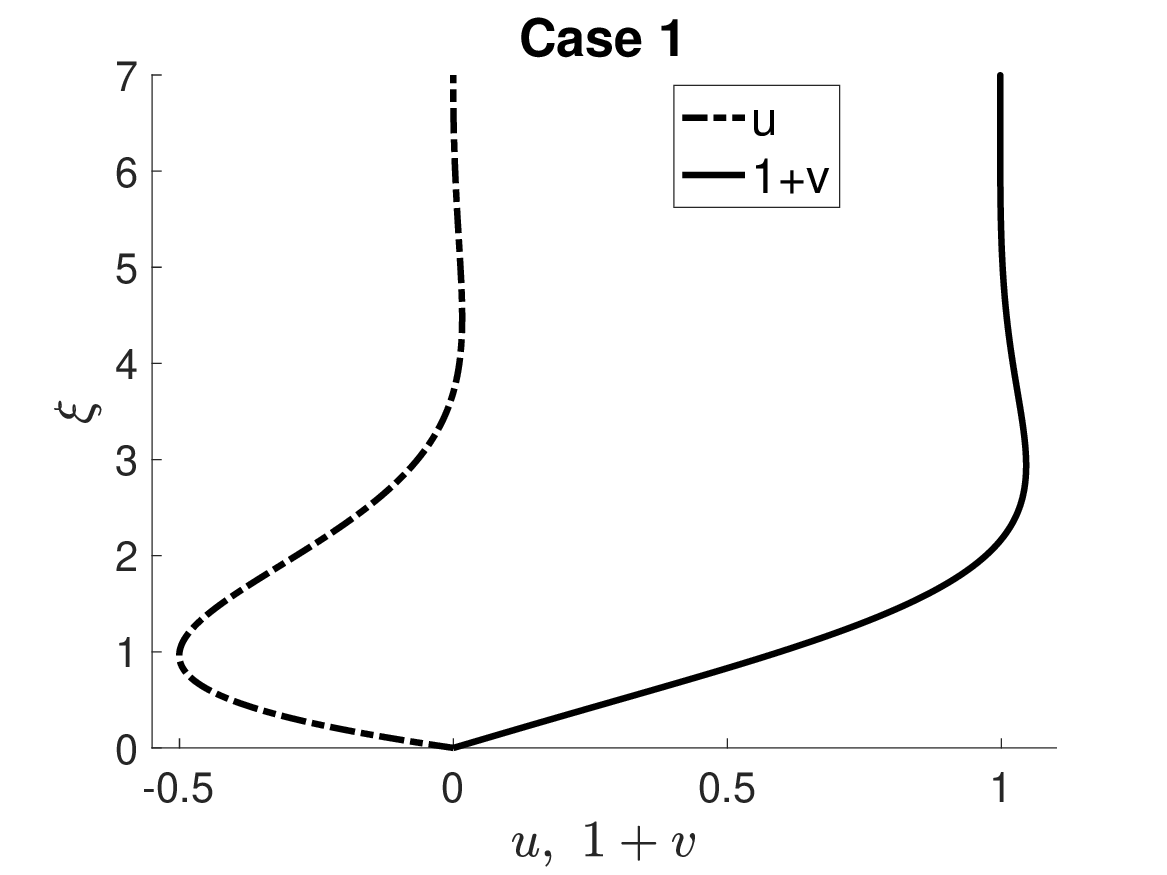}
        \caption*{(a)}  
    \end{subfigure}%
    \hfill  
    \begin{subfigure}[b]{0.49\textwidth}  
        \centering
        \includegraphics[width=\textwidth]{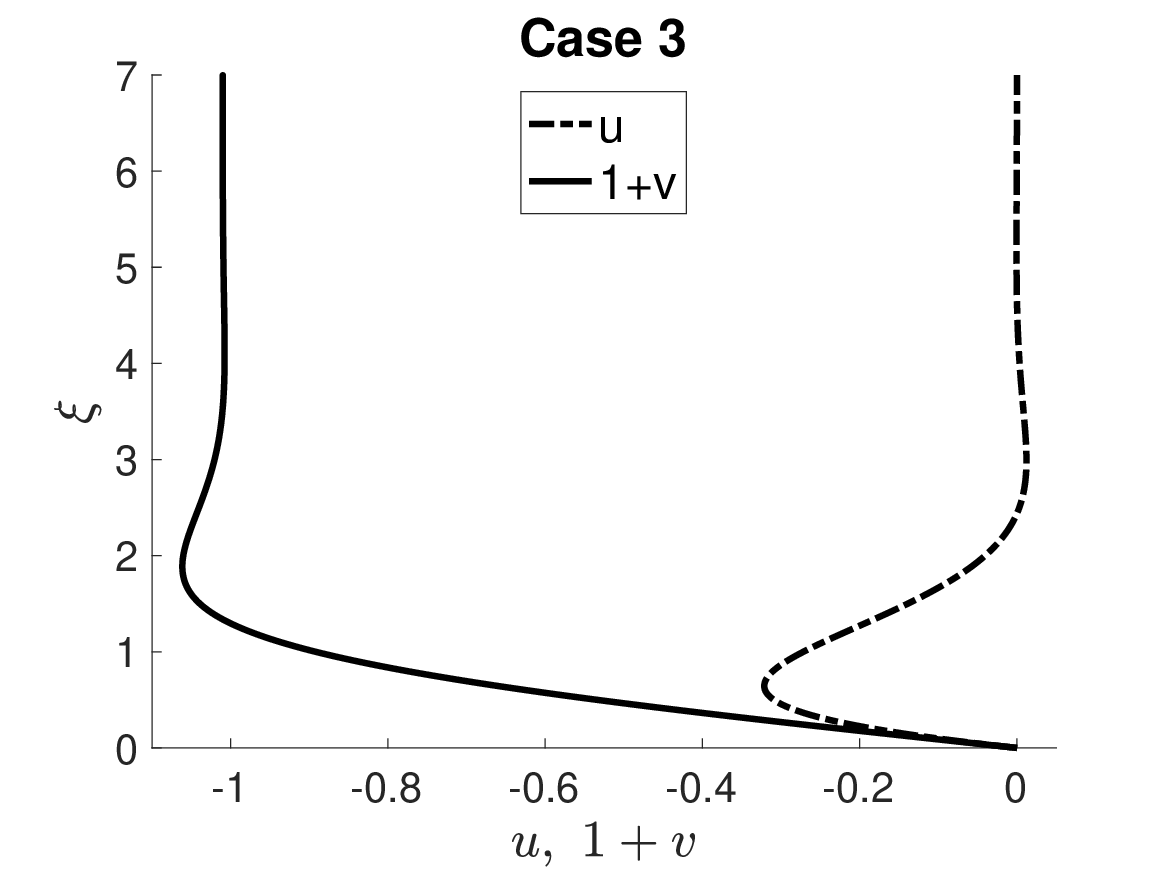}
        \caption*{(b)}  
    \end{subfigure}
    \caption{Numerical solution for Cases 1 and 3 of Table~\ref{tab:n_cases}.}
    \label{fig:n_cases_numSoln}
\end{figure}

We expect velocity profiles to resemble those of the classical Ekman layer solution. 
$u$ should feature a jet with negative peak velocity near the surface and decay further aloft; $1+v$ should be positive and be characterized by an elevated maximum and asymptote to a value of $1$ aloft to match the gradient wind. 

We see that the expected behaviour is followed in Case 1 but not Case 3. 
For Case 3, $1+v$ does not tend to $1$ aloft. 
The stable numerical solution for $1+v$ appears like a mirror image about $0$ of what we would have expected based on the Ekman layer solution. 

We find that there is a switch in the stable numerical solution from a form similar to Case 1 (Fig.~\ref{fig:n_cases_numSoln},a) where $v$ has a positive slope near the ground, to that seen in Case 3 (Fig.~\ref{fig:n_cases_numSoln},b) where $v$ has a negative slope near the ground. 
Through some trial and error, we find that this switch occurs at $n \approx 0.5$, with a mild sensitivity to $Ro$.
(Note that a Neumann boundary condition $du/d\xi = dv/d\xi = 0$ is used for Case 3.
Using a Dirichlet condition instead would have caused a sharp change in $v$ aloft to match the boundary condition.)
The effect on the profiles upon changing $Ro$ is minimal (not shown).
 
To analyse possible solutions to \eqref{eq:nonDimmedEqns1} and \eqref{eq:nonDimmedEqns2} aloft, we set $d^2u/d\xi^2\ = d^2v/d\xi^2 = 0$ (the effects of curvature in the velocity profiles should only be present close to the ground). 
We get two algebraic equations
\begin{equation}\label{eq:algebraicEqns}
    u^2 + v^2 + \left(2 + \frac{1}{Ro}\right)v = 0 \ , \qquad \qquad
    uv + u\left((1-n) + \frac{1}{Ro}\right) = 0\ .
\end{equation}
There are four solutions to this system of algebraic equations, which are
\begin{equation}\label{eq:algebraicSolns}
\begin{gathered}
    u = 0,\ v = 0\ , \qquad \qquad u = 0,\ v = -\left(2 + \frac{1}{Ro}\right)\ , \\ \\
    u = \pm\sqrt{\left((1-n) + \frac{1}{Ro}\right)(1+n)},\ v = -\left((1-n) + \frac{1}{Ro}\right) \ .
\end{gathered}
\end{equation}
In Fig.~\ref{fig:n_cases_numSoln}(b), we see that the $u\to0$ and $(1+v)\to-\left(1 + 1/Ro\right)$. 
This is consistent with a possible solution in \eqref{eq:algebraicSolns}, and therefore appears to be a mathematically valid tendency. 
But this is clearly not physical, since the velocity should approach that of the gradient wind aloft.

One simple reason for this switch could be the increasing dominance of the nonlinear terms as $n$ becomes larger. 
This is seen clearly from \eqref{eq:nonDimmedSimple1} and \eqref{eq:nonDimmedSimple2}. 
For even a moderately large $Ro$ (say, $Ro > 10$), the effect of $1/Ro$ can be neglected. 
Focusing on $n$, we see that as $n$ becomes larger, the coefficient attached to $u$ in \eqref{eq:nonDimmedSimple2} is much smaller than that attached to $uv$, and hence the nonlinear behaviour dominates the linear. 

Details of the numerical solver are given in Appendix \ref{section:appendixNumSoln} and a qualitative mechanism for this flip is described. 
Additionally, in Appendix \ref{section:appendixNumSoln}, we plot velocity profiles with a vertically varying eddy viscosity (Fig.~\ref{fig:appendix_varEddViscShearStress},a). 
Even with the variable eddy viscosity, the switch of solution behaviour is seen. 
However, the $n$ at which this switch occurs appears to be pushed to a higher value ($n \approx 0.67$ for the chosen eddy viscosity profile).

\citet{meng1995analytical}, \citet{kepert2001dynamics} and other subsequent work \citep{snaiki2017modeling,fang2018novel, yang2021height} consider a slip condition instead at the bottom boundary.
The switch in solution behaviour for large $n$ is observed with this boundary condition as well. 
For example, using $C = 0.002$ and $K = 50\, \mathrm{m^2s^{-1}}$ as in \citet{kepert2001dynamics}, the switch is observed at $n\approx 0.69$ (Fig.~\ref{fig:appendix_varEddViscShearStress},b). Raising $C$ to a larger value, such as $0.02$, causes the switch to occur at a lower $n\approx0.55$.

The observed behaviour limits the physical validity of the governing equations and suggests that care must be taken when using large values of $n$ to ensure that the solution to the equations retains physical tendencies. 

\section{Approximate analytical solution}\label{section:analyticalSoln}
Acknowledging the above limitations of the governing equations, we constrain the study in this section to $n\in [0, 0.45]$. 
Increases in the parameters $Ro$ and $n$ make the solution increasingly dependent on the nonlinear terms.
In all the cases considered, the coefficient attached to the nonlinear terms are of the same order of magnitude as those of the linear terms, and therefore nonlinear terms cannot be neglected.

We follow Lyapunov's artificial small parameter method \citep{nayfeh2008perturbation}. 
Attaching an artificial small parameter $\delta$ to the nonlinear terms, the equations are
\begin{eqnarray}\label{eq:artificialSmallParam1}
\frac{d^2 u}{d \xi^2} &=& 
- \tilde{\alpha} v
- \delta\tilde{\gamma}(u^2 + v^2) \ ,
\\
\label{eq:artificialSmallParam2}
\frac{d^2 v}{d \xi^2} & = & 
\tilde{\beta} u+
\delta\tilde{\gamma} uv \ .
\end{eqnarray}
Consider a series expansion of the form
\begin{eqnarray}\label{eq:seriesExpansion}
        u(\xi) = & u_0(\xi) + \delta u_1(\xi)+ \delta^2 u_2(\xi) + \cdots \ , \\ 
        v(\xi) = & v_0(\xi) + \delta v_1(\xi)+ \delta^2 v_2(\xi) + \cdots \ .
\end{eqnarray}
Substituting the series and collecting orders of $\delta$, we obtain the following systems of equations at the first few orders. 
The leading order equations read
 \begin{eqnarray}\label{eq:orderEqn_0_a}
 \frac{\partial^2 u_0}{\partial \xi^2} &=&  -\tilde{\alpha} v_0 \ , \\
 \label{eq:orderEqn_0_b}
\frac{\partial^2 v_0}{\partial \xi^2}  &=&  
\tilde{\beta}u_0 \ , 
\end{eqnarray}
with boundary conditions $u_0(0) = u_0(\infty) =
v_0(\infty) = 0$, $v_0(0) = -1$. 
Similarly, the first order correction reads
\begin{eqnarray}\label{eq:orderEqn_1_a}
    \frac{\partial^2 u_1}{\partial \xi^2}&=& -\tilde{\alpha}v_1 - \tilde{\gamma}(u_0^2 + v_0^2) \ , \\
    \label{eq:orderEqn_1_b}
\frac{\partial^2 v_1}{\partial \xi^2}&=&\tilde{\beta}u_1 + \tilde{\gamma}(u_0v_0) \ ,
\end{eqnarray}
with boundary conditions $u_1(0) = u_1(\infty) =
v_1(0) = v_1(\infty) = 0$.
The second order correction is
\begin{eqnarray}\label{eq:orderEqn_2_a}
    \frac{\partial^2 u_2}{\partial \xi^2} &=&  -\tilde{\alpha}v_2 - \tilde{\gamma}(2u_0u_1 + 2v_0v_1)
 \ , \\
\label{eq:orderEqn_2_b}
\frac{\partial^2 v_2}{\partial \xi^2}  &=&  
\tilde{\beta}u_2 + \tilde{\gamma}(u_0v_1 + u_1v_0) \ ,
\end{eqnarray}
with the same boundary conditions as for the first order problem.

Note that we have used the no-slip boundary condition instead of a slip condition. This choice is made to come up with a sufficiently simple form of the solution including nonlinear effects that can represent the various trends seen in more complicated analyses. Future versions of this analysis that include a varying eddy viscosity or a higher fidelity boundary condition would likely match realistic velocity profiles even more closely.

The solution to the equations at the leading order is simply that of the Ekman layer, namely
\begin{eqnarray}\label{eq:orderSoln_0u}
    u_0 &=& -\frac{\tilde{\alpha}}{2}{\mathrm{e}}^{-\xi}\sin(\xi) \ , \\
    \label{eq:orderSoln_0v}
    v_0 &=& -{\mathrm{e}}^{-\xi}\cos(\xi) \ .
\end{eqnarray}   
To obtain the first-order correction, substitute for $u_0$ and $v_0$ from Eqns.~\eqref{eq:orderSoln_0u} and \eqref{eq:orderSoln_0v} into \eqref{eq:orderEqn_1_b} and \eqref{eq:orderEqn_1_a}, and then substitute $u_1$ from \eqref{eq:orderEqn_1_b} in \eqref{eq:orderEqn_1_a}. The resulting equation can be solved by considering an ansatz which is a linear combination of terms in the following product of sets
\begin{equation}\label{eq:ansatz}
    \left\{1,\ e^{-\xi},\ e^{-2\xi}\right\}\times
    \left\{1,\ \cos(\xi),
    \ \sin(\xi),\ \cos(2\xi),
    \ \sin(2\xi)
    \right\} \ .
\end{equation}
Substituting the ansatz, comparing terms and solving for the coefficients, we obtain
\begin{eqnarray}\label{eq:orderSoln_1u}
\begin{aligned}
    u_1(\xi) = 
    \frac{\tilde{\gamma}}{10}\left( 
    \frac{\tilde{\alpha}^2}{4} + 1
    \right)e^{-\xi}\cos(\xi) + 
    \frac{\tilde{\gamma}}{30}e^{-\xi}\sin(\xi) \\
-
\frac{\tilde{\gamma}}{10}\left( 
    \frac{\tilde{\alpha}^2}{4} + 1
       \right)e^{-2\xi} + 
       \frac{\tilde{\gamma}}{30}\left( 
    -\frac{3}{8}\tilde{\alpha}^2 +2
       \right)e^{-2\xi}\sin(2\xi) \ ,
\end{aligned}\\ \label{eq:orderSoln_1v}
       \begin{aligned}
       v_1(\xi) = 
    \frac{\tilde{\gamma}}{15\tilde{\alpha}}e^{-\xi}\cos(\xi) - 
    \frac{\tilde{\gamma}}{5\tilde{\alpha}}\left(\frac{\tilde{\alpha}^2}{4} + 1\right)e^{-\xi}\sin(\xi)\\
- \frac{\tilde{\gamma}}{10\tilde{\alpha}}\left(\frac{\tilde{\alpha}^2}{4} + 1\right)e^{-2\xi} +
\frac{\tilde{\gamma}}{30\tilde{\alpha}}\left(\frac{3\tilde{\alpha}^2}{4} + 1\right)e^{-2\xi}\cos(2\xi) \ .
       \end{aligned} 
\end{eqnarray}
The solutions beyond the first order become increasingly unwieldy and difficult to interpret conceptually. 
Further, as elaborated below, the first-order solution already represents a noticeable enhancement in accuracy compared to the linear leading-order solution, and is also sufficiently close to the numerical solution.
This work will therefore focus primarily on the first-order solution.  
Appendix~\ref{section:appendixhigherOrderSolns} presents the second order correction for the interested reader. 

Noting that the gradient wind vector is seldom constant with height in the HBL, Appendix~\ref{section:appendixlinVarGradWind} generalizes the above solution to linearly decreasing gradient winds.
\subsection{Comparison with numerical solution}
\begin{figure}[htpb]
     \centering
     \begin{subfigure}{0.49\textwidth}
         \centering
    \includegraphics[width=\textwidth]{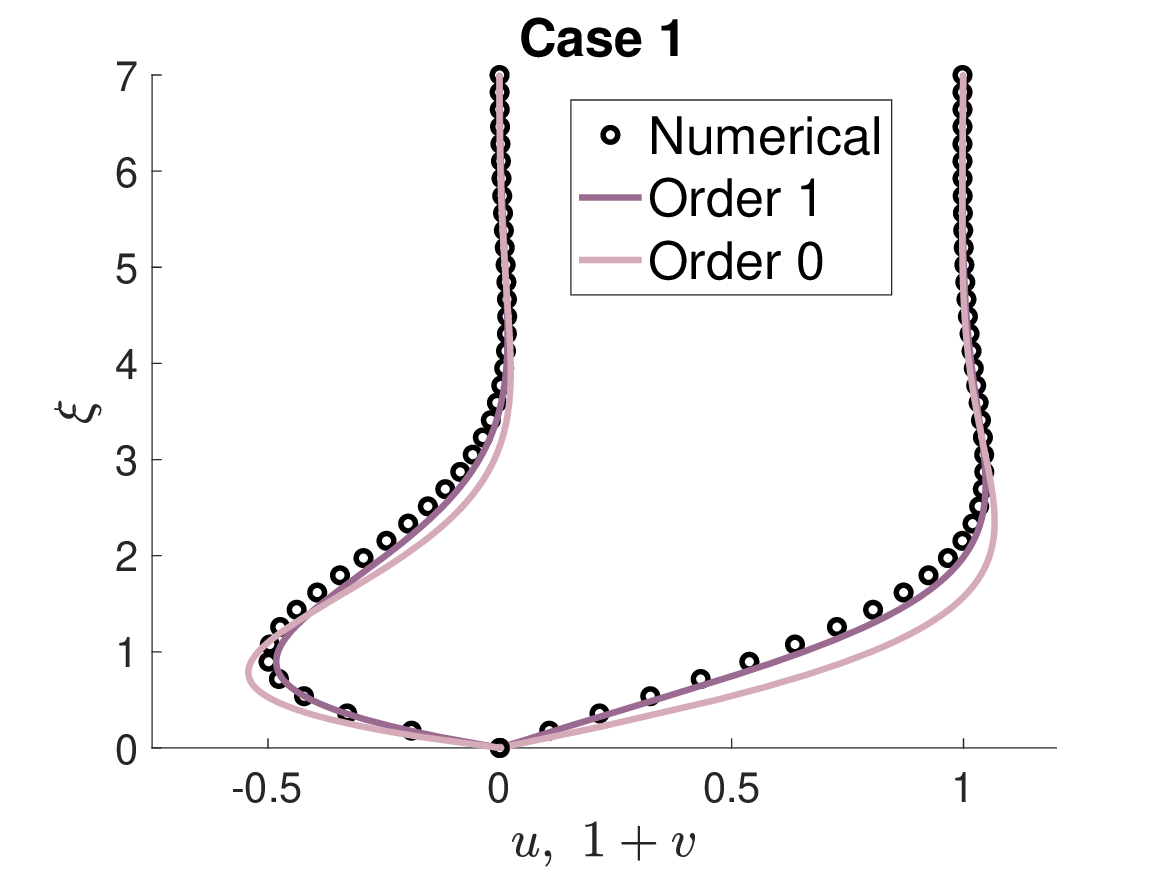}
    \caption*{(a)}
     \end{subfigure}
     \begin{subfigure}{0.49\textwidth}
         \centering
        \includegraphics[width=\textwidth]{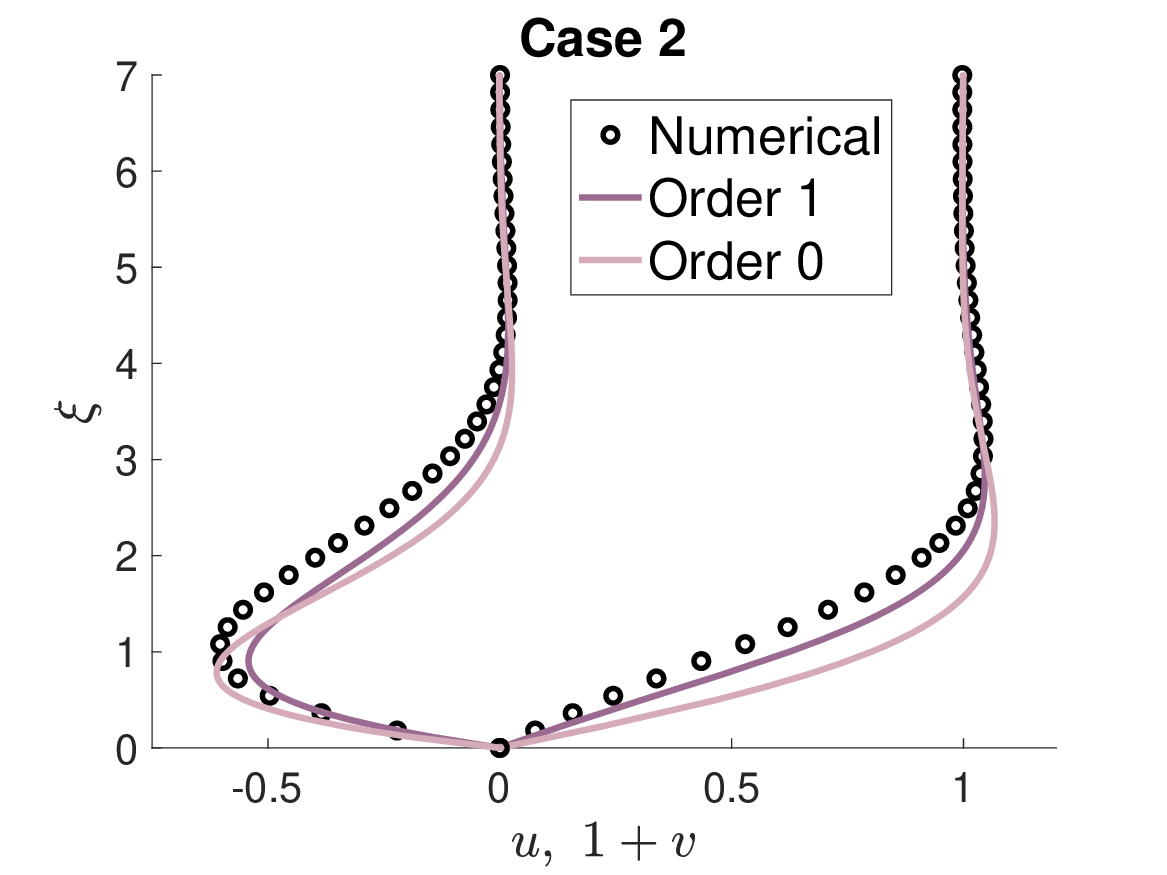}
        \caption*{(b)}
     \end{subfigure}
     \caption{\label{fig:n_cases_anlnumSoln}Analytical and numerical solution for cases 1 and 2 of Table~\ref{tab:n_cases}.}
\end{figure}

Leading and first order solutions are contrasted against the numerical solution in Fig.~\ref{fig:n_cases_anlnumSoln} for cases 1 and 2 of Table~\ref{tab:n_cases}. 
Case 1 has a small $n = 0.3$ that falls well within the solution's region of validity. Based on the figure, it is apparent
that a first-order correction is sufficient to closely match the numerical solution.
The corrections modulate the inflow, yielding slightly more diffused inflow and tangential velocity profiles.
Case 2 is characterized by a relatively larger $n = 0.45$, which is close to the bordering $n \approx 0.5$ beyond which the switch in stable solution was observed (\S~\ref{section:physValidity}). 
The solution features a slower convergence rate, requiring a larger number of terms to obtain an adequate approximation.
However, these higher order solutions are required only when $n$ becomes quite close to (say, within $0.05$ of) the critical value, and a first order solution remains sufficient for smaller $n$. 
For $n\gtrapprox 0.5$, the analytical solution diverges.

The correction term $v_1$ is negative near the ground (upto $\xi\approx 3)$ and $u_1$ has an S-like shape where it is positive at the ground and negative above. 
Both the $u_1$ and $v_1$ terms thus have a tendency to reduce the magnitudes of the peak radial and tangential velocities respectively.

The role of the corrections in moderating the gradient of the velocity near the ground is also interesting. 
The derivative of $v_1$ is negative upto $\xi\approx 1$, and thus decreases the slope of tangential velocity at the ground.
As $n$ increases, the strength of this reduction in the slope of $v_0$ by $v_1$ also increases. 
This may be connected to the switch in solution behaviour, and is elaborated in Appendix~\ref{section:appendixNumSoln}.
Overall, the proposed corrections offer improvements to the linear (leading-order) solution and additional insight into the nonlinear effects.

\section{Sensitivity Analysis}\label{section:sensAnalysis}
To characterize the role of the input parameters, we analyse here the sensitivity of the normalized solutions to their variations.
A discussion on sensitivities is also presented for the solutions in the more intuitive dimensional form.

\begin{figure}[htpb]
     \centering
     \begin{subfigure}{0.49\textwidth}
         \centering
    \includegraphics[width=\textwidth]{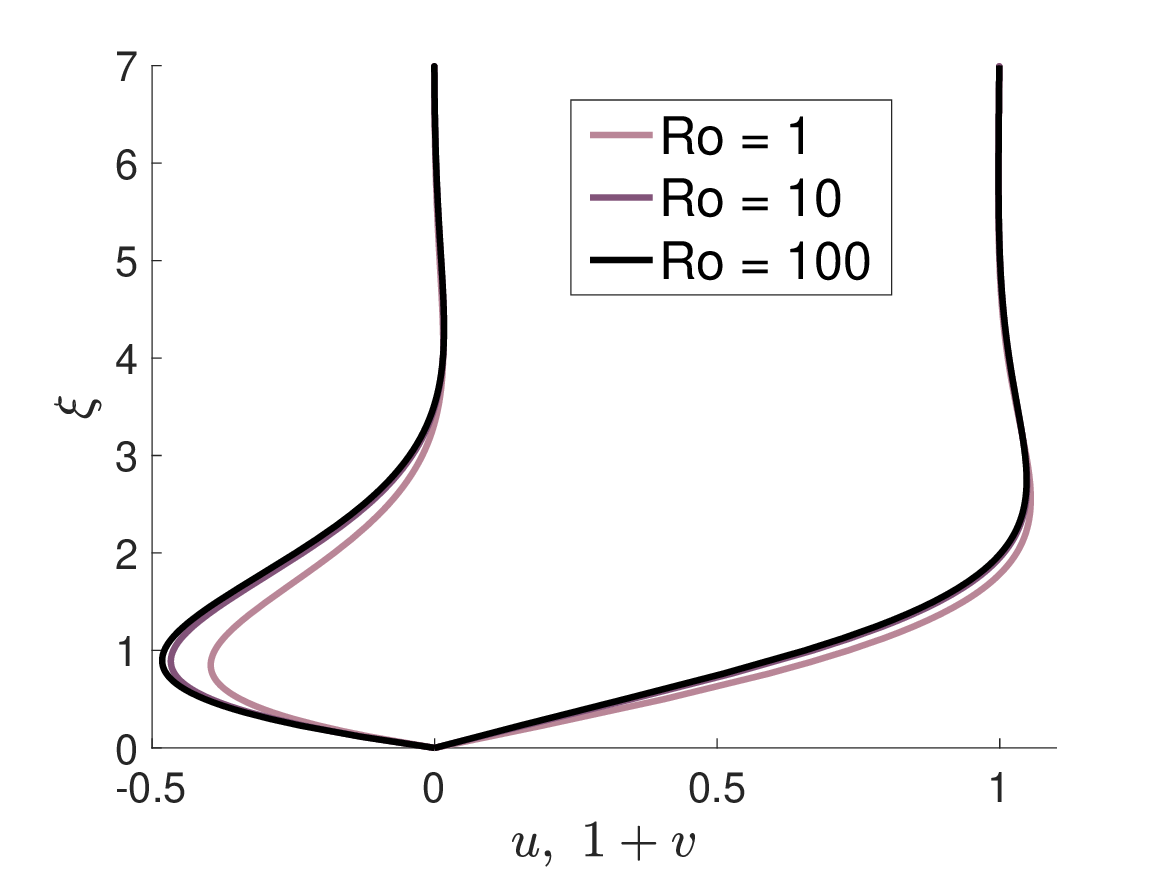}
         \caption*{(a)}
     \end{subfigure}
     \begin{subfigure}{0.49\textwidth}
         \centering
        \includegraphics[width=\textwidth]{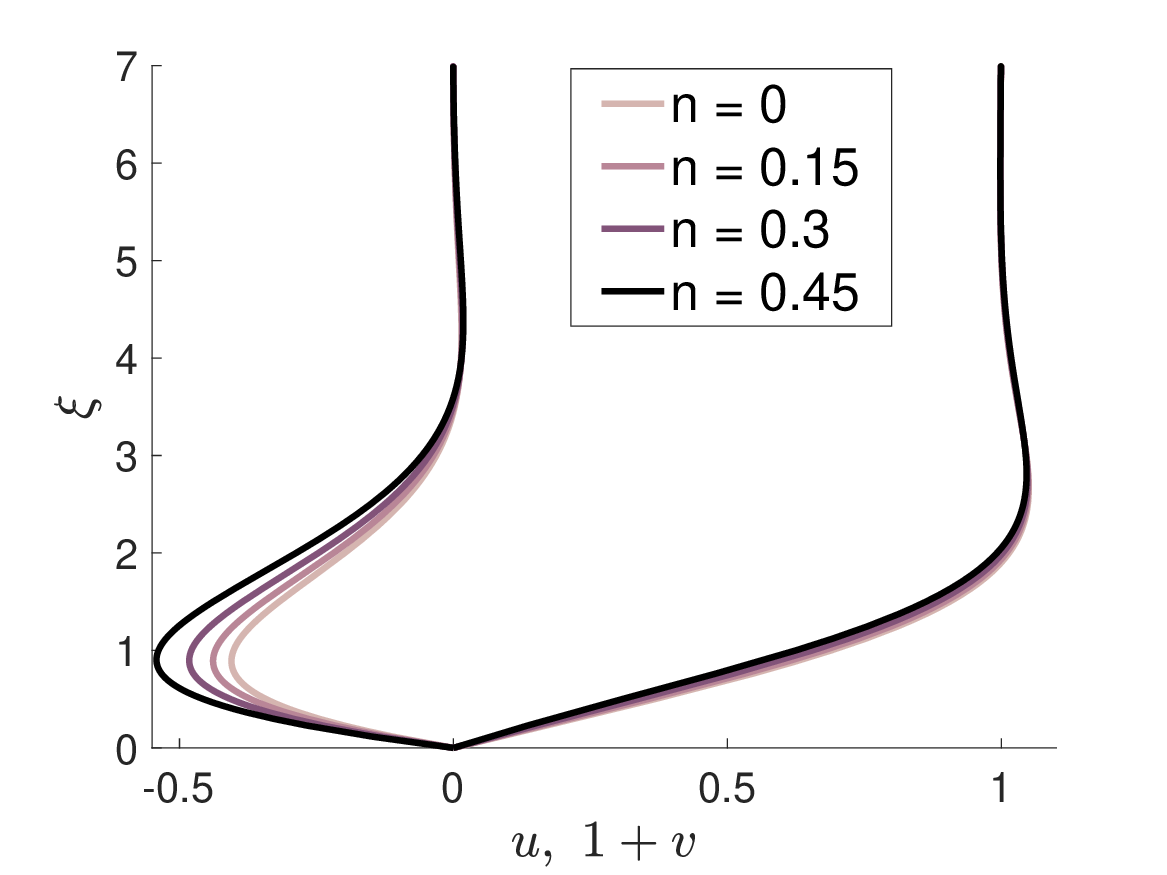}
         \caption*{(b)}
     \end{subfigure}
     \caption{\label{fig:Ro_n_sensitivity}Sensitivity of the first order accurate solutions in normalized form to $Ro$ and $n$.}
\end{figure}

Figure \ref{fig:Ro_n_sensitivity}(a) shows the sensitivity to $Ro$ for $n = 0.3$. 
For a given latitude, a small $Ro$ means either that $G$ is small or that $R$ is large. 
Thus, small $Ro$ is not a very interesting case. 
When $Ro$ becomes large, we expect the effect of the latitude to be negligibly small. Wind profiles should get closer to each other as $Ro$ increases. 
This is reflected in Fig.~\ref{fig:Ro_n_sensitivity}(a), with the curve for $Ro = 10$ appearing almost superimposed on the curve for $Ro = 100$. 
Figure \ref{fig:Ro_n_sensitivity}(b) plots the sensitivity to $n$ for $Ro = 100$. 
As $n$ increases, so does the strength of the inflow. 
The sensitivity to $n$ for $1+v$ is, however, far less pronounced than for $u$. 

Sensitivities of the solutions in dimensional form are discussed next. 
Solutions are dimensionalized based on parameter values listed in Table~\ref{tab:dimValues}, with the exception of the parameter that is being varied. 
\begingroup
    \setlength{\tabcolsep}{12pt}
\renewcommand{\arraystretch}{1.5}
\begin{table}
    \centering
    \begin{tabular}{ccccc}
    \hline
    \hline
        $\mathbf{G}$ & $40$  m s\textsuperscript{-1} & &
$\mathbf{n}$ & 0.25  \\
$\mathbf{K}$ & $50$  m\textsuperscript{2}s\textsuperscript{-1} & &
 $\mathbf{R}$ & $50$  km\\
$\mathbf{f}$ & $5\times10^{-5}$ s\textsuperscript{-1} & &
  $\mathbf{G_z}$ & $0$ s\textsuperscript{-1} \\
        \hline
        \hline
    \end{tabular}
    \caption{Dimensional values of parameters used to plot Figures~\ref{fig:dim_Gn_sens} to \ref{fig:dim_fGz_sens}.}
    \label{tab:dimValues}
    \end{table}
\begin{figure}[htpb]
     \centering
     \begin{subfigure}{0.49\textwidth}
         \centering
    \includegraphics[width=\textwidth]{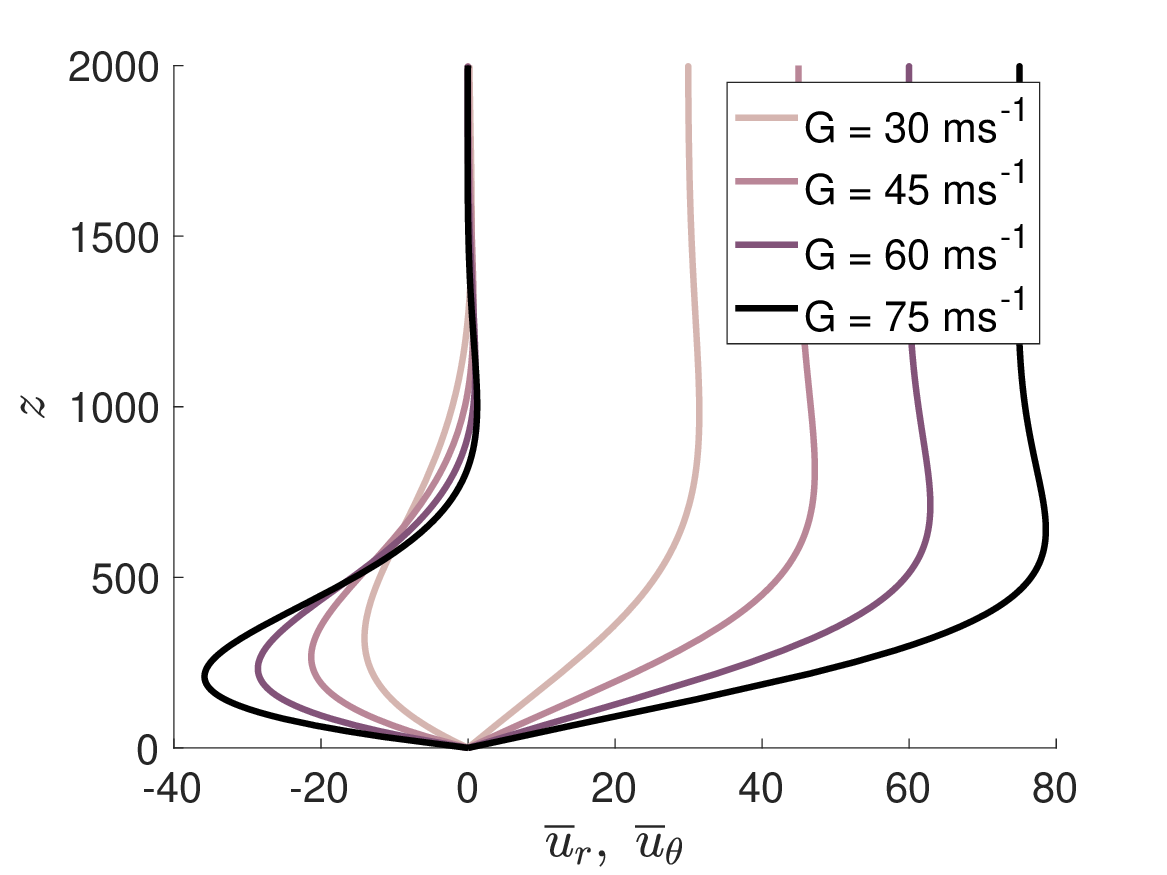}
         \caption*{(a)}
     \end{subfigure}
     \begin{subfigure}{0.49\textwidth}
         \centering
        \includegraphics[width=\textwidth]{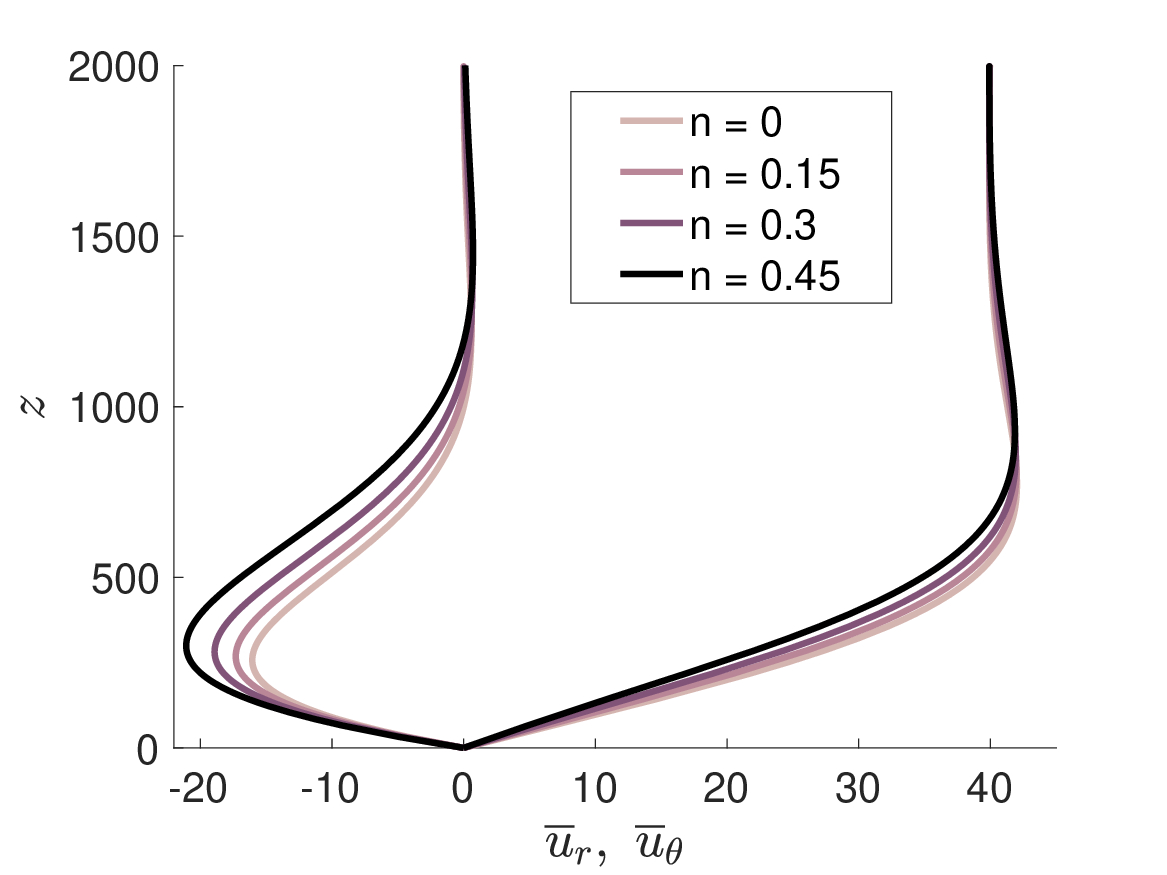}
         \caption*{(b)}
     \end{subfigure}
\caption{\label{fig:dim_Gn_sens}Sensitivity of the first order accurate solution in dimensional form to $G$ and $n$. $z$ is in metres and the velocity components in metres per second.}
\end{figure}
Figure \ref{fig:dim_Gn_sens}(a) shows the sensitivity to $G$. 
As is expected, $G$ has a strong impact on the shape and magnitude of the velocity components. 
As $G$ increases, the $\overline{u}_\theta$ profiles feature a strengthened jet with a more pronounced nose moving closer to the surface. A similar behavior characterizes the $\overline{u}_r$ profile. 
The sensitivity to variations in $n$ is shown in Fig. \ref{fig:dim_Gn_sens}(b). 
As was seen in the non-dimensional analysis in Fig.\ref{fig:Ro_n_sensitivity}(b), despite its importance in determining the convergence of the analytical solution, this parameter only modulates the $\overline{u}_r$ profile with a modest impact on $\overline{u}_\theta$.
The sensitivity to $K$ is shown in Fig.~\ref{fig:dim_Rnu0_sens}(a).
As $K$ increases, there is increased mixing, causing the jet to become wider. 
With larger $K$, the peak of this more diffuse jet also tends to become higher. 
The role of $K$ here is similar to that in the Ekman layer.
The effect of increasing $R$ (Fig.~\ref{fig:dim_Rnu0_sens},b) is quite similar to that of increasing $K$.
As $R$ increases, the jet becomes less pronounced and thicker, and its elevation above the surface level increases. 
This is consistent with the intuition that as we move farther away from the eye of the hurricane, the effect of the centrifugal force term in the gradient wind forcing becomes smaller, and the wind profile tends to that due to a geostrophic forcing.
The mild decrease in the peak magnitudes of $\overline{u}_r$ and $\overline{u}_\theta$ with increasing $R$ is due to the change in $Ro$ and this effect is negligible as seen in the figure.
\begin{figure}[htpb]
     \centering
     \begin{subfigure}{0.49\textwidth}
         \centering
    \includegraphics[width=\textwidth]{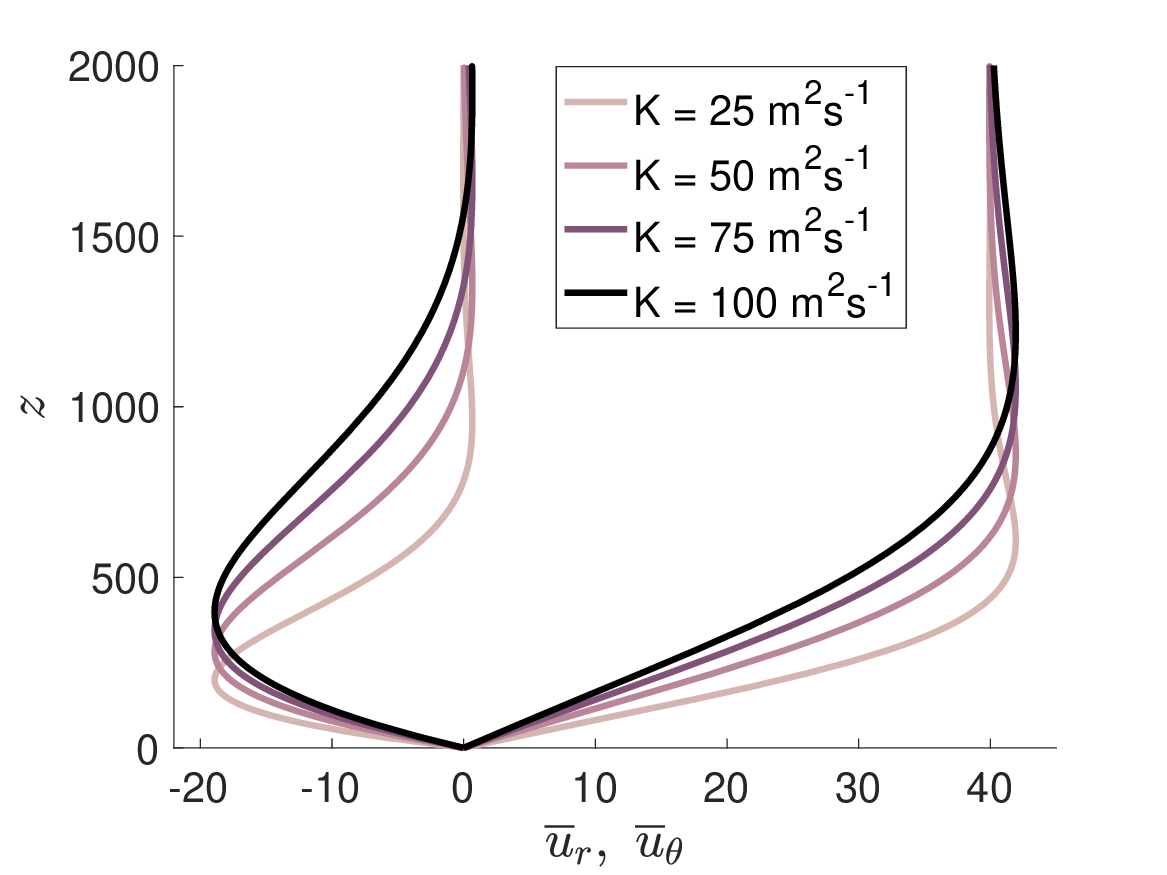}
         \caption*{(a)}
     \end{subfigure}
     \begin{subfigure}{0.49\textwidth}
         \centering
        \includegraphics[width=\textwidth]{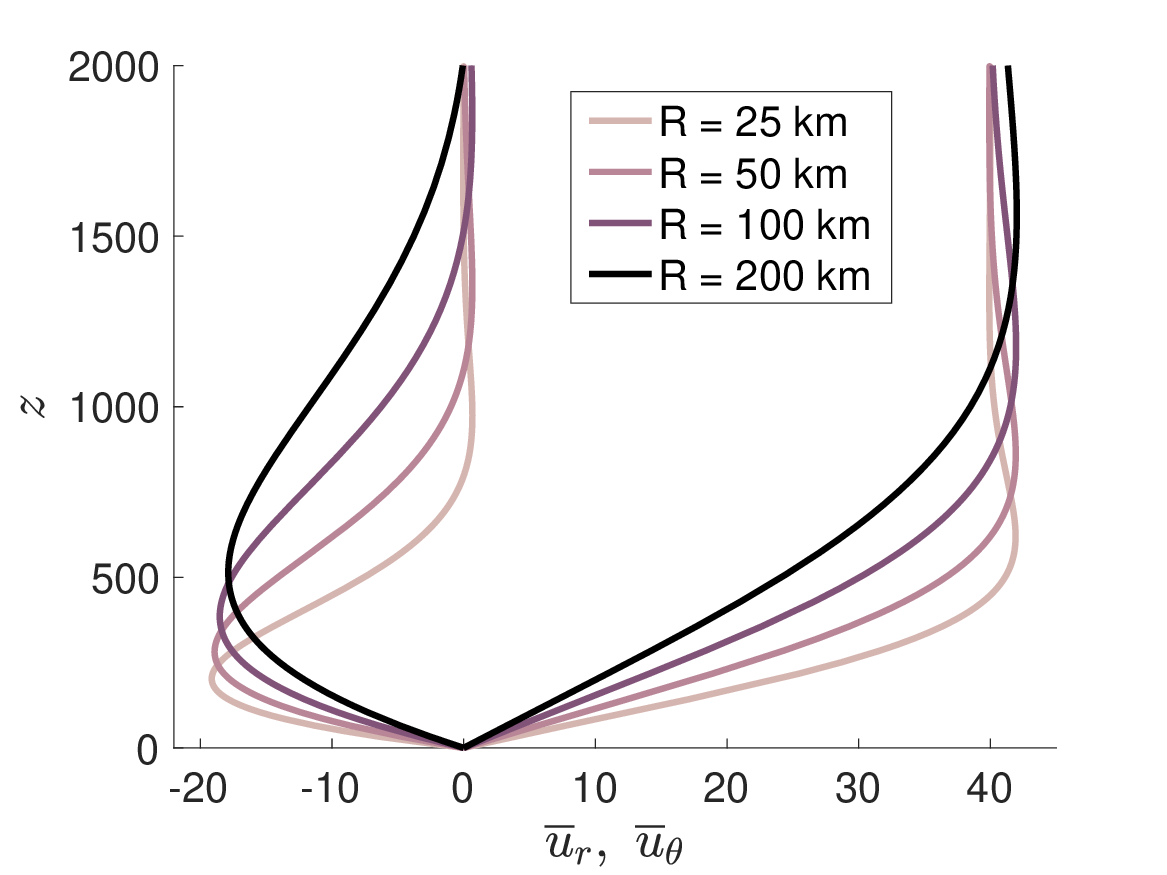}
         \caption*{(b)}
     \end{subfigure}
\caption{\label{fig:dim_Rnu0_sens}Sensitivity to $K$ and $R$.}
\end{figure}

These sensitivities can be further analysed quantitatively using the analytical solution. 
For example, we examine the peak magnitude and height of the inflow and use the leading order solution for the radial velocity for simplicity.
 Differentiating \ref{eq:orderSoln_0u} and setting to zero, one obtains $\xi_{\overline{u}_r}^\text{peak} = \pi/4$, corresponding to
\begin{equation}\label{eq:nonlininflowPeakDim}
\begin{gathered}
      z_{\overline{u}_r}^\text{peak} = 1.11\sqrt{\frac{KR}{G}}\left(
    \left(\frac{1}{Ro}+2\right)
    \left(\frac{1}{Ro}+(1-n)\right)
    \right)^{-0.25}
    \approx 0.93\sqrt{\frac{KR}{G\sqrt{1-n}}} \, .
\end{gathered}
\end{equation}
where in the last step we have use $1/Ro\approx 0$. 
From this formula we see that, to the leading order, $R$ and $K$ have the same effect on the height of the inflow peak. 
This was also observed in the two parts of Fig.~\ref{fig:dim_Rnu0_sens}.
From the formula, as $G$ and $n$ increase, the height of the peak decreases, which is also seen in Fig.~\ref{fig:dim_Gn_sens}. Substituting in \ref{eq:orderSoln_0u}, the magnitude of peak inflow is 
\begin{equation}\label{eq:inflowPeakMag}
    \overline{u}_r^\text{peak} = 0.32G\sqrt{\frac{\frac{1}{Ro}+2}{\frac{1}{Ro}+(1-n)}} \approx 0.46\frac{G}{\sqrt{1-n}} \, .
\end{equation}
This formula shows that as $n$ increases, the magnitude of peak inflow increases. 
This is seen in Fig.~\ref{fig:dim_Gn_sens}(b). 
Figure \ref{fig:dim_fGz_sens}(a) shows the sensitivity to $f$.
Coriolis frequency has a small impact for this radius, since the centrifugal term dominates the Coriolis term. 
The impact of $f$ is slightly more visible for a larger radius of, say $R = 200\, \mathrm{km}$ (not shown). 
But for cases of interest, i.e., for small $R$ and large $G$, the effect due to $f$ is negligible.

The sensitivity to variations in $G_z$ is shown in Fig.~\ref{fig:dim_fGz_sens}(b). 
As apparent, this parameter mainly affects the slope of $\overline{u}_\theta$ away from the surface, with minimal impact on the $\overline{u}_r$ profiles.

\begin{figure}[htpb]
     \centering
     \begin{subfigure}{0.49\textwidth}
         \centering
    \includegraphics[width=\textwidth]{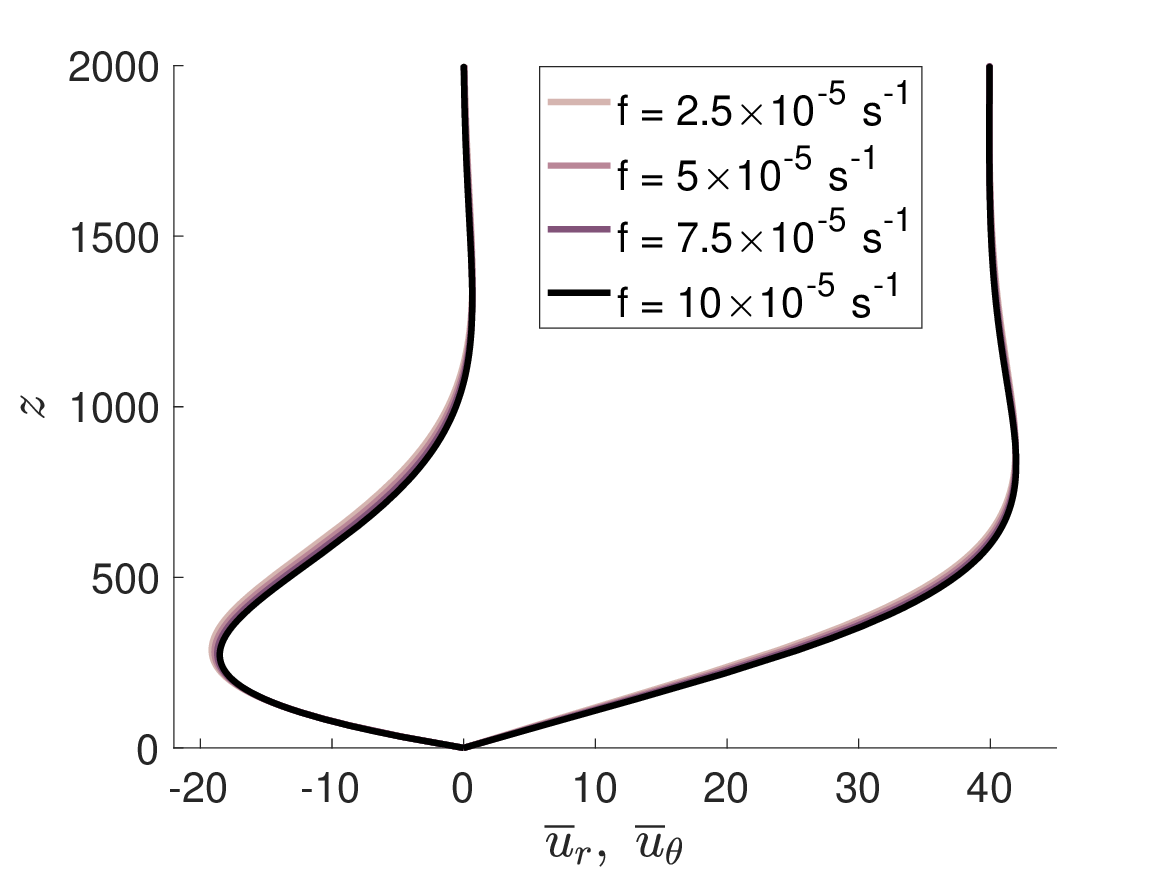}
         \caption*{(a)}
     \end{subfigure}
     \begin{subfigure}{0.49\textwidth}
         \centering
        \includegraphics[width=\textwidth]{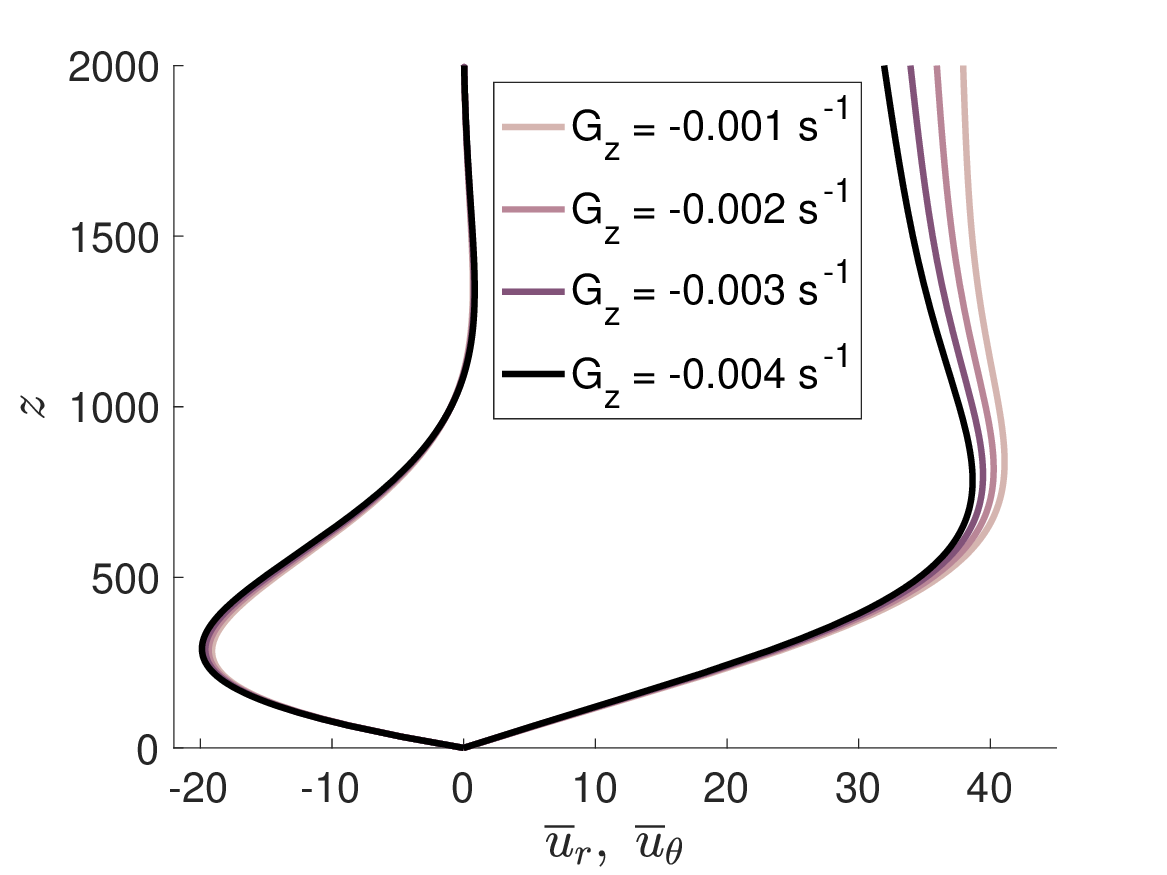}
         \caption*{(b)}
     \end{subfigure}
\caption{\label{fig:dim_fGz_sens}Sensitivity to $f$ and $G_z$.}
\end{figure}
Although \eqref{eq:inflowPeakDim} was derived using the linear solution \eqref{eq:orderSoln_0u}, we can extend this to the nonlinear solution using the fact the solution is insensitive to $Ro$ (see Fig.~\ref{fig:Ro_n_sensitivity},a). 
Neglecting $1/Ro$, the solution to \eqref{eq:nonDimmedEqns1} and \eqref{eq:nonDimmedEqns2} is only a function of $\xi$ and $n$.
For a fixed $n$, the non-dimensional peak is a constant value.
In a dimensional setting, this thus means that even the nonlinear solution features
\begin{equation}\label{eq:inflowPeakDim}
      z_{\overline{u}_r}^\text{peak} \ , 
      z_{\overline{u}_\theta}^\text{peak}
    \propto \sqrt{\frac{KR}{G}}f(n) \, .
\end{equation}
If we vary only $G$ (or, alternatively, $K$ or $R$) holding the other constants fixed, then height of the peak velocity values for the leading and first order analytical and numerical solutions are simply multiples of each other. 
We demonstrate this in Fig.~\ref{fig:nonlinearSens}, where normalized variations of peak heights for a range of parameter sweeps are shown.
Normalized peak heights are defined as 
\begin{equation}
    \zeta_G^{\overline{u}_i} = \frac{z_{\overline{u}_i}^\text{peak}(G,\boldsymbol{q}_0)-z_{\overline{u}_i}^\text{peak}(G_0,\boldsymbol{q}_0)}{z_{\overline{u}_i}^\text{peak}(G_0,\boldsymbol{q}_0)} \, ,
\end{equation}
where $\overline{u}_i$ denotes either $\overline{u}_r$ or $\overline{u}_\theta$, and $\boldsymbol{q}_0$ refers to the other dimensional parameters that we hold fixed.
We see that the six curves overlap for variations in $G$, $R$ and $K$.
The overlap is weaker for larger $G$ and $R$, due to the small effect from $Ro$.

The above reasoning does not however hold for $n$, and this is confirmed by Fig.\ref{fig:nonlinearSens}(b) where no collapse is observed.
We also note that the numerical solution shows a faster growth of peak height with $n$ than the linear solution.
The first order analytical solution is an improvement to the linear solution but fails to match the numerical solution as $n$ increases.
The figure also shows that, for the numerical and first order solutions, the rate of growth of the peak height is higher for tangential velocity than radial velocity. 
This is not captured by the linear solution, for which the dashed curve overlaps the solid one.
\begin{figure}[htpb]
     \centering
     \begin{subfigure}{0.45\textwidth}
         \centering
    \includegraphics[width=\textwidth]{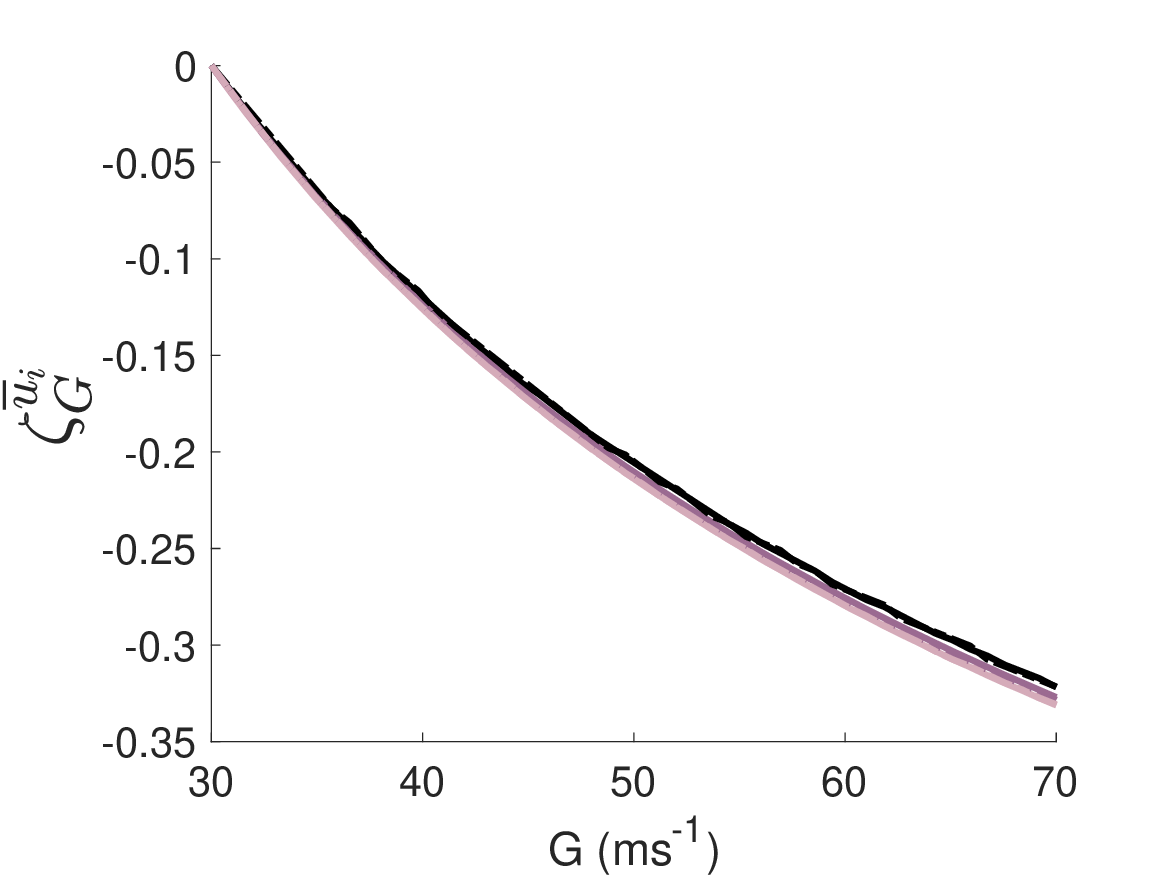}
         \caption*{(a)}
     \end{subfigure}%
     \begin{subfigure}{0.45\textwidth}
         \centering
        \includegraphics[width=\textwidth]{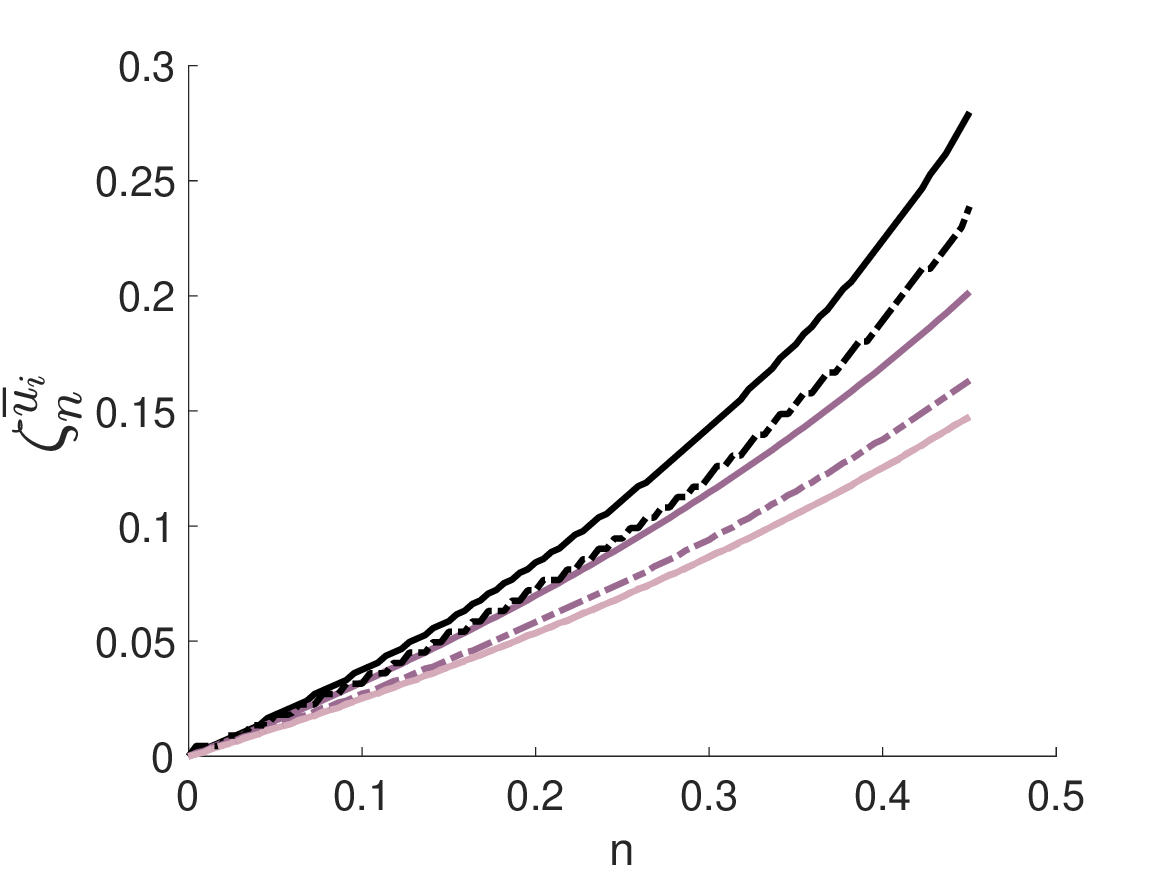}
         \caption*{(b)}
     \end{subfigure}
     \begin{subfigure}{0.45\textwidth}
         \centering
        \includegraphics[width=\textwidth]{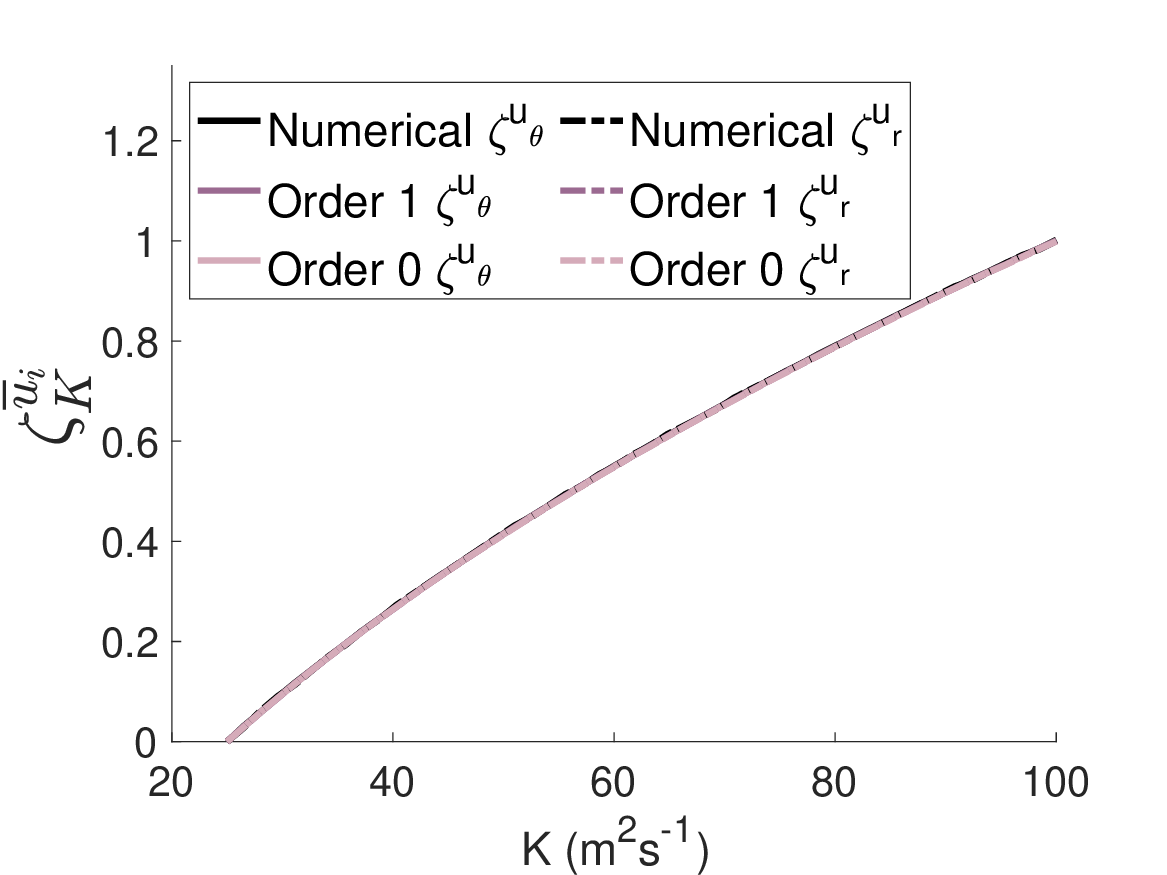}
         \caption*{(c)}
     \end{subfigure}%
     \begin{subfigure}{0.45\textwidth}
         \centering
        \includegraphics[width=\textwidth]{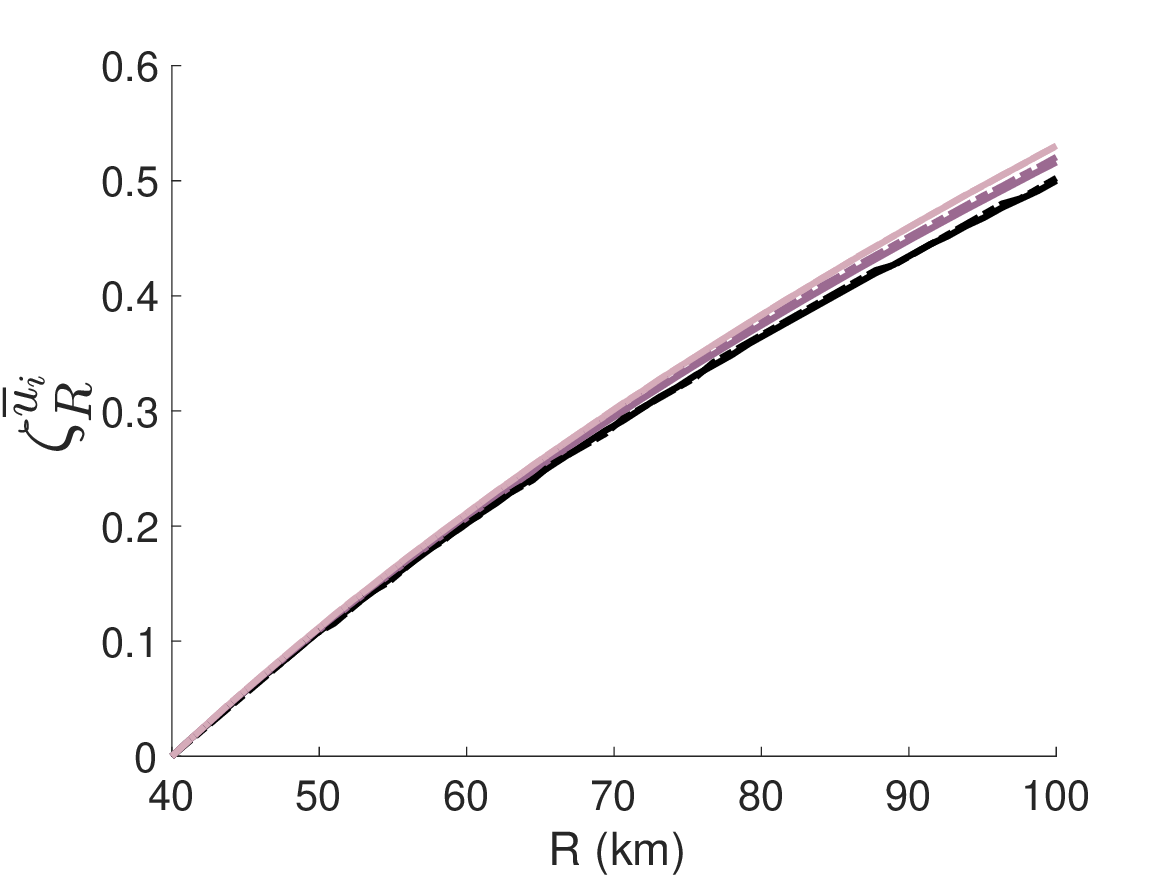}
         \caption*{(d)}
     \end{subfigure}
    \caption{\label{fig:nonlinearSens} Normalized variations of peak heights for the tangential and radial velocities against variations of selected dimensional parameters.}
\end{figure}
\section{Velocity Magnitude}\label{section:velMag}
Figure \ref{fig:VelMagNewStrategy}(a) plots the numerical solution for the velocity magnitude $\sqrt{u^2 + (1+v)^2}$, for a range of representative $Ro$ and $n$ values. 
As apparent from the figure, the solution is insensitive to both of these parameters. 
This is surprising, especially when considering the role of $n$ in controlling the convergence of the analytical solution.

\begin{figure}[htpb]
     \centering
     \begin{subfigure}{0.49\textwidth}
         \centering
    \includegraphics[width=\textwidth]{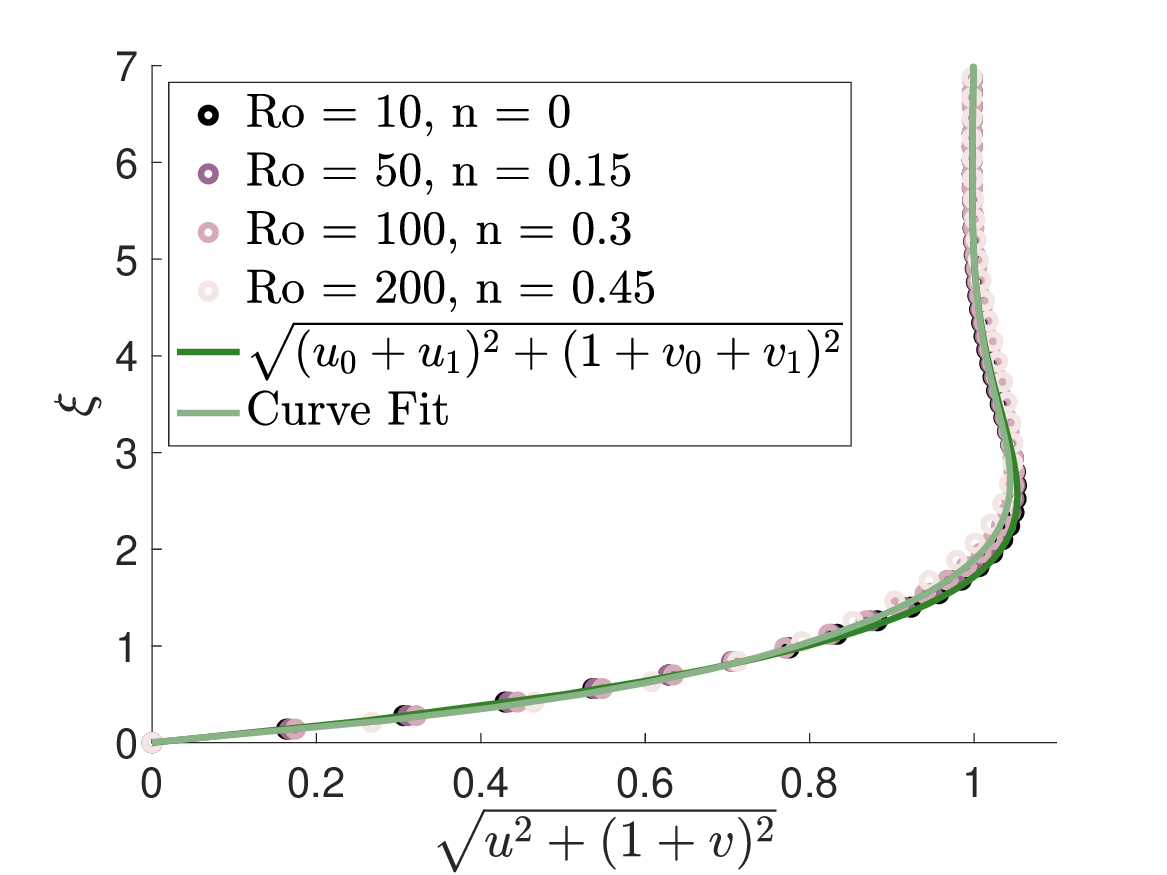}
    \caption*{(a)}
     \end{subfigure}
     \begin{subfigure}{0.49\textwidth}
         \centering
        \includegraphics[width=\textwidth]{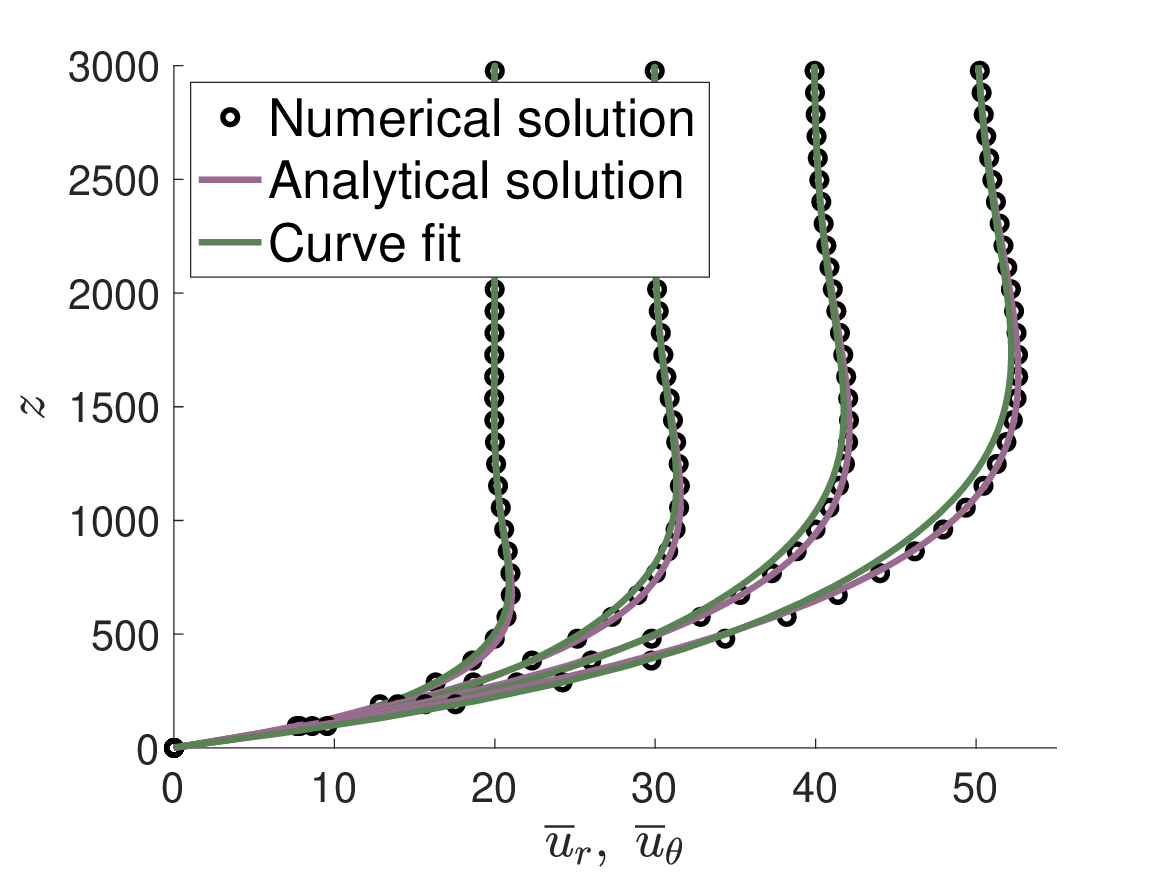}
         \caption*{(b)}
     \end{subfigure}
\caption{\label{fig:VelMagNewStrategy}Left: Numerical solution for the velocity magnitude in normalized form, for various $n$ and $Ro$. Figure also includes the analytical solution~\ref{eq:velMagAnalyticalSoln_u},~\ref{eq:velMagAnalyticalSoln_v} and the curve fit \ref{eq:curveFit}. Right: Numerical, analytical and curve fit solutions in dimensional form. The curves from left to right are cases 1 to 4 of Table~\ref{tab:Table3} respectively.}
\end{figure}
If one is interested only in the magnitude of velocity, then this finding presents an alternative strategy to obtain the velocity profile for larger $n$. 
We had noted in Sect.~\ref{section:analyticalSoln} that for small $n$, the analytical solution with first order correction is indistinguishable from the numerical solution. 
We can therefore consider the case when $Ro\to\infty$ and $n = 0$ for the first order analytical solution. 
In this limit, the coefficients are constant numerical values. Substituting $1/Ro = n = 0$ in \eqref{eq:orderSoln_0u}, \eqref{eq:orderSoln_0v}, \eqref{eq:orderSoln_1u} and \eqref{eq:orderSoln_1v}, we obtain
\begin{eqnarray}\label{eq:velMagAnalyticalSoln_u}
\begin{aligned}
        u_0 + u_1 = 
        -1.37e^{-\xi}\sin(\xi) + 
        0.42e^{-\xi}\cos(\xi) \\
        - 0.42e^{-2\xi}
        - 0.05e^{-2\xi}\sin(2\xi) \ ,
\end{aligned}\\
        \label{eq:velMagAnalyticalSoln_v}
        \begin{aligned}
        1 + v_0 + v_1 = 
        1-0.97e^{-\xi}\cos(\xi)
        -0.30e^{-\xi}\sin(\xi) \\
        - 0.15e^{-2\xi}
        + 0.12e^{-2\xi}\cos(2\xi)  \ .
        \end{aligned}
\end{eqnarray}

Since the coefficients are just numerical values (and not functions of $Ro$ or $n$), an even simpler strategy could be to perform a curve fit. 
A simple fit $C(\xi)$ of the velocity magnitude yields
\begin{equation}\label{eq:curveFit}
\begin{split}
    C(\xi) = 1 -0.88e^{-\xi}\cos(\xi) -0.28e^{-\xi}\sin(\xi) \\
    -0.13e^{-2\xi}\cos(2\xi)  + 0.31e^{-2\xi}\sin(2\xi) \ .
\end{split}
\end{equation}
Figure~\ref{fig:VelMagNewStrategy}(a) also includes $\sqrt{(u_0 + u_1) + (1 + v_0 + v_1)^2}$ from \eqref{eq:velMagAnalyticalSoln_u} and~\eqref{eq:velMagAnalyticalSoln_v}, and the curve fit $C(\xi)$ in \eqref{eq:curveFit}. 
We see that both of these are quite close to the other four curves.
This solution can also be tested for robustness in a dimensional setting. 
We consider four curves, with various values of the dimensional parameters listed in Table \ref{tab:Table3}. 
The black symbols in Fig.~\ref{fig:VelMagNewStrategy}(b) are obtained by solving the dimensional equations \eqref{eq:toBeSolvedDimensional1} and \eqref{eq:toBeSolvedDimensional2} numerically with boundary conditions \eqref{eq:dimensionalBCs}. 
The purple curves labelled ``analytical" are \begin{equation}\label{eq:simpleAnalytical}
    G\sqrt{\left(u_0\left(\frac{z}{H}\right) + u_1\left(\frac{z}{H}\right)\right)^2 + \left(1+v_0\left(\frac{z}{H}\right) + v_1\left(\frac{z}{H}\right)\right)^2} \ ,
\end{equation}
where $u_0+u_1$ are $1+v_0+v_1$ are from~\eqref{eq:velMagAnalyticalSoln_u} and~\eqref{eq:velMagAnalyticalSoln_v}. 
The green curves are $GC(z/H)$ where $C(\xi)$ is from~\eqref{eq:curveFit}.
We see that curves from the analytical solution and the curve fit are able to match the numerical solution very closely. 
The curve fit shows a slight deviation for larger $n$ (see rightmost curve), but the analytical solution superposes the numerical solution almost exactly. 
This proves further that the simpler strategies derived in this section are robust alternatives to the series expansion from Sect.\ref{section:analyticalSoln}, when considering the velocity magnitude.

\begingroup
    \setlength{\tabcolsep}{12pt}
\renewcommand{\arraystretch}{1.5}
\begin{table}
    \centering
    \begin{tabular}{cccccc}
    \hline
         & $\mathbf{G}$ (m~s\textsuperscript{-1}) & $\mathbf{R}$ (km) & $\mathbf{f}$ (s\textsuperscript{-1}) & $\mathbf{n}$ & $\mathbf{K}$ (m\textsuperscript{2}~s\textsuperscript{-1}) \\
         \hline
        \textbf{Case 1} & $20$ & $50$ & $2.5\times10^{-5}$ &
        0.1 & 10 \\
        \textbf{Case 2} & $30$ & $100$ & $5\times10^{-5}$ &
        0.2 & 20 \\
        \textbf{Case 3} & $40$ & $150$ & $7.5\times10^{-5}$ &
        0.3 & 30 \\
        \textbf{Case 4} & $50$ & $200$ & $10\times10^{-5}$ &
        0.4 & 40 \\
        \hline
    \end{tabular}
    \caption{Parameters used in the four curves of Fig.~\ref{fig:VelMagNewStrategy}(b).}
    \label{tab:Table3}
\end{table}
\endgroup

Using this new analytical solution, we can now obtain the depth and peak value of the velocity magnitude by differentiating $\sqrt{(u_0 + u_1) + (1 + v_0 + v_1)^2}$ and setting to 0.
\begin{equation}\label{eq:velMagPeakLoc}
    \xi_{\text{peak}} = 2.56 \implies z_{\text{peak}} = 3.62\sqrt{\frac{KR}{G}}\left(
    \left(\frac{1}{Ro}+2\right)
    \left(\frac{1}{Ro}+(1-n)\right)
    \right)^{-0.25} \ .
\end{equation}
For large Rossby number, this simplifies further to
\begin{equation}\label{eq:peakHeight}
    z_{\text{peak}} = 3.04\sqrt{\frac{KR}{G\sqrt{1-n}}} \ .
\end{equation}
The dependence of the height of the peak from \ref{eq:peakHeight} is also seen in Figs.~\ref{fig:dim_Gn_sens} and~\ref{fig:dim_Rnu0_sens}. 
As $K$ and $R$ increase, the height of the peak tangential velocity increases. 
Whereas as $G$ increases, the height of the peak decreases slightly. 
The sensitivity of the depth appears to be greater for $K$ and $R$ because they have a much broader range of physical values than does $G$. 
The coefficient 3.04 is larger than the coefficient 0.93 obtained earlier, indicating that the peak inflow occurs at a lower height than that of the velocity magnitude. This is consistent with the solution to the Ekman layer.

Substituting~\ref{eq:peakHeight} in ~\ref{eq:simpleAnalytical} gives us a maximum dimensional velocity magnitude of $1.0528 \, G$. 
Notice that this value depends only on the gradient wind magnitude. 
Thus, while the height at which the peak occurs is dependent on the viscosity, radius etc., the percentage supergradient predicted by this analysis, i.e. $5.28\%$, is a constant. 
The existing literature largely focuses on the radius of maximum wind (RMW) within the eyewall, where much higher supergradient values are reported \citep[e.g.,][]{kepert2001nonlinear}. 
We emphasize that our analysis is constrained to the region outside the eyewall. 
Despite this different focus, the value we obtain, $5.28\%$, still falls within the range of supergradient values reported in some of these studies \citep[e.g.,][]{yang2021height}.

\section{Summary and Conclusions}\label{section:summary}
This work has presented approximate solutions to the nonlinear equations describing the vertical structure of the horizontal velocity profiles in the HBL under constant and linearly varying gradient wind. 
The formulation is based on a series-expansion approach and it has been shown that including a first-order correction to the linear leading order solution increases the accuracy of the model while retaining a relatively simple form.
The solution has been compared against a linear counterpart and a corresponding ``exact'' numerical solution across a range of realistic values of the input parameters in both normalized and dimensional form.
Velocities were scaled with gradient wind speed $G$ and vertical coordinate $z$ was scaled with $\sqrt{2K/I}$, following \citet{kepert2001dynamics}. 
We found that $\sqrt{2K/I}$ remains a robust choice of scaling for $z$ even when accounting for the nonlinear terms in the governing equations, further corroborating findings from the numerical study of \citet{foster2009boundary}. 

It was shown that nonlinear terms cause a switch in the stable solution from a physical one, where $\overline{u}_\theta\to G$ aloft, to a nonphysical one where $\overline{u}_\theta\to -G\left(1 + 1/Ro\right)$.
Thus, care must be taken when prescribing $n$, so that the solution to the equations retains physical tendencies. 
This also prompts a closer look at the modeling assumption $\partial v/\partial r\approx -nV_g/R$. 
A better approximation to this radial derivative as a function of $z$ (for example, $\partial v/\partial r\approx -nv/R$) instead of a constant or linear function throughout the height of the boundary layer could yield an approximation that is closer to reality.

From a wind profiles perspective, the non-dimensional first order solution exhibits a reduced peak velocity magnitude and more elevated jets when compared to its leading-order counterpart.
In normalized form, the parameter $Ro$ was found to have a negligible impact on the velocity profiles. 
In dimensional form, the solution was found, as expected, to be sensitive to variations in $G$, $R$ and $K$ and insensitive to variations in $f$. 
These were used to show that both the height of inflow and tangential velocity peak in both the linear and nonlinear solutions follow a $\sqrt{KR/G}$ dependence.
The parameter $n$, along with its role in controlling the convergence of the analytical solution, was found to primarily control the inflow profile. 
It however does not have a significant effect on the tangential velocity and velocity magnitude. 
Building on this lack of sensitivity to variations in $Ro$ and $n$, a simple alternative expression was proposed to describe the velocity magnitude profile. 
This simplified expression provides insight into the dependence of the jet height and magnitude on the various parameters and shows that the peak jet magnitude is $5.28\%$ supergradient.

In summary, the proposed series solution provides a nuanced yet still conceptually simple description of velocity profiles in the HBL and associated sensitivites, accounting for the impact of non-linear terms.
Although valid only away from the eyewall and under the assumption of constant eddy viscosity, this formulation serves as a foundation for developing operational models that capture the correct sensitivities of the hurricane boundary layer profile.

\vfill

\section{Acknowledgements}
This research was funded by the National Institute of Standards and Technology (\url{ror.org/05xpvk416}) under Grant No. 70NANB22H057. 
A special thanks to Dr. Jaeyoung Jung, Postdoctoral Scholar, Columbia University for his help with the numerical ODE solver. 

\bibliographystyle{spbasic_updated.bst} 
\bibliography{arxivRefs.bib}

\section{Appendices}
\subsection{Appendix 1: Governing Equations}\label{section:appendixGovEqns}
The Navier Stokes equations in cylindrical coordinates, in a reference frame rotating with the Earth and moving with the hurricane are
\begin{equation} \label{eq:appendix_mom_r} \frac{\partial u_r}{\partial t} + u_r\frac{\partial u_r}{\partial r} + \frac{u_\theta}{r}\frac{\partial u_r}{\partial \theta} + u_z\frac{\partial u_r}{\partial z} - \frac{u^2_\theta}{r} = -\frac{1}{\rho}\frac{\partial P}{\partial r} + fu_\theta + D_{
\nu_m,r
} \ ,
\end{equation}
\begin{equation} \label{eq:appendix_mom_theta}  \frac{\partial u_\theta}{\partial t} + u_r\frac{\partial u_\theta}{\partial r} + \frac{u_\theta}{r}\frac{\partial u_\theta}{\partial \theta} + u_z\frac{\partial u_\theta}{\partial z} + \frac{u_ru_\theta}{r} = -\frac{1}{r\rho}\frac{\partial P}{\partial \theta} - fu_r + D_{
\nu_m,\theta} \ , \end{equation}
\begin{equation} \label{eq:appendix_mom_z} \frac{\partial u_z}{\partial t} + u_r\frac{\partial u_z}{\partial r} + \frac{u_\theta}{r}\frac{\partial u_z}{\partial \theta} + u_z\frac{\partial u_z}{\partial z}  = -\frac{1}{\rho}\frac{\partial P}{\partial z} + D_{\nu_m,z} \end{equation}
where the terms $D_{\nu_m}$ represent molecular diffusion.

The continuity equation in cylindrical coordinates is
\begin{equation} \label{eq:appendix_continuity}
    \frac{\partial u_r}{\partial r} + \frac{u_r}{r} + \frac{1}{r}\frac{\partial u_\theta}{\partial \theta} + \frac{\partial u_z}{\partial z} = 0 \ .
\end{equation}
Multiplying the continuity equation \eqref{eq:appendix_continuity} by $u_r$ and adding to the radial momentum equation \eqref{eq:appendix_mom_r}, by $u_\theta$ and adding to \eqref{eq:appendix_mom_theta}, and by $u_z$ and adding to \eqref{eq:appendix_mom_z}, we obtain
\begin{equation} \label{eq:appendix_mom_r_cons} \frac{\partial u_r}{\partial t} + \frac{\partial (u_ru_r)}{\partial r} + 
\frac{1}{r}\frac{\partial (u_ru_\theta)}{\partial \theta} + 
\frac{\partial (u_ru_z)}{\partial z} + \frac{u^2_r}{r} = 
-\frac{1}{\rho}\frac{\partial P}{\partial r} 
+ \frac{u^2_\theta}{r} + fu_\theta  
+ D_{\nu_m,r} \ ,
\end{equation}
\begin{equation} \label{eq:appendix_mom_theta_cons}
\frac{\partial u_\theta}{\partial t} 
+ \frac{\partial (u_ru_\theta)}{\partial r} 
+ \frac{1}{r}\frac{\partial (u_\theta u_\theta)}{\partial \theta} 
+ \frac{\partial (u_zu_\theta)}{\partial z} + \frac{u_ru_\theta}{r}= -\frac{1}{r\rho}\frac{\partial P}{\partial \theta} 
- \frac{u_ru_\theta}{r} - fu_r 
+ D_{\nu_m,\theta} \ , 
\end{equation}
\begin{equation} \label{eq:appendix_mom_z_cons}
\frac{\partial u_z}{\partial t} + \frac{\partial (u_ru_z)}{\partial r} + \frac{1}{r}\frac{\partial (u_\theta u_z)}{\partial \theta} + \frac{\partial (u_zu_z)}{\partial z} + \frac{u_ru_z}{r} = -\frac{1}{\rho}\frac{\partial P}{\partial z} + D_{
\nu_m,z} \ . 
\end{equation}
Taking the time-average, we have
\begin{equation} \label{eq:appendix_mom_r_timeavg}
\frac{\partial \overline{(u_ru_r)}}{\partial r} + 
\frac{1}{r}\frac{\partial \overline{(u_ru_\theta)}}{\partial \theta} + 
\frac{\partial \overline{(u_ru_z)}}{\partial z} 
+ \frac{\overline{u_r^2}}{r}= 
-\frac{1}{\rho}\frac{\partial \overline{P}}{\partial r} 
+ \frac{\overline{u^2_\theta}}{r} + f\overline{u}_\theta  
+ D_{\nu_m,r} \ ,
\end{equation}
\begin{equation} \label{eq:appendix_mom_theta_timeavg}
\frac{\partial \overline{(u_ru_\theta)}}{\partial r} 
+ \frac{1}{r}\frac{\partial \overline{(u_\theta u_\theta)}}{\partial \theta} 
+ \frac{\partial \overline{(u_zu_\theta)}}{\partial z} 
+ \frac{\overline{u_ru_\theta}}{r}
= -\frac{1}{r\rho}\frac{\partial \overline{P}}{\partial \theta} 
- \frac{\overline{u_ru_\theta}}{r} - f\overline{u}_r 
+ D_{\nu_m,\theta} \ , 
\end{equation}
\begin{equation} \label{eq:appendix_mom_z_timeavg}
\frac{\partial \overline{(u_ru_z)}}{\partial r} 
+ \frac{1}{r}\frac{\partial \overline{(u_\theta u_z)}}{\partial \theta} 
+ \frac{\partial \overline{(u_zu_z)}}{\partial z}  
+ \frac{\overline{u_ru_z}}{r}= -\frac{1}{\rho}\frac{\partial \overline{P}}{\partial z} + D_{
\nu_m,z} \ . 
\end{equation}
The vertical momentum equation does not hold much physical significance far from the eyewall, so it can be ignored for this analysis.

The gradient wind balance is expected to hold in the radial direction. Also, no average pressure gradient is expected in the azimuthal direction. That is,
\begin{eqnarray}\label{eq:appendix_gradWind}
        \frac{1}{\rho}\frac{\partial \overline{P}}{\partial r} = \frac{V_g^2}{r} + fV_g \ , \\
        \label{eq:appendix_gradWind_theta}
        \frac{1}{\rho}\frac{\partial \overline{P}}{\partial \theta} = 0 \ .
\end{eqnarray}
Substituting, we get
\begin{equation} \label{eq:appendix_r_subgradwind}
\frac{\partial \overline{(u_ru_r)}}{\partial r} + 
\frac{1}{r}\frac{\partial \overline{(u_ru_\theta)}}{\partial \theta} + 
\frac{\partial \overline{(u_ru_z)}}{\partial z} 
+ \frac{\overline{u_r^2}}{r}= 
-\left(\frac{V_g^2}{r} + fV_g\right) 
+ \frac{\overline{u^2_\theta}}{r} + f\overline{u}_\theta  
+ D_{\nu_m,r} \ ,
\end{equation}
\begin{equation} \label{eq:appendix_theta_subgradwind}
\frac{\partial \overline{(u_ru_\theta)}}{\partial r} 
+ \frac{1}{r}\frac{\partial \overline{(u_\theta u_\theta)}}{\partial \theta} 
+ \frac{\partial \overline{(u_zu_\theta)}}{\partial z} 
+ \frac{\overline{u_ru_\theta}}{r}
= 
- \frac{\overline{u_ru_\theta}}{r} - f\overline{u}_r 
+ D_{\nu_m,\theta} \ .
\end{equation}
The azimuthal direction can be considered statistically homogeneous. 
Hence
\begin{equation} \label{eq:appendix_r_azimuthalHomo}
\frac{\partial \overline{(u_ru_r)}}{\partial r} + 
\frac{\partial \overline{(u_ru_z)}}{\partial z} 
+ \frac{\overline{u_r^2}}{r}= 
-\left(\frac{V_g^2}{r} + fV_g\right) 
+ \frac{\overline{u^2_\theta}}{r} + f\overline{u}_\theta  
+ D_{\nu_m,r} \ ,
\end{equation}
\begin{equation} \label{eq:appendix_theta_azimuthalHomo}
\frac{\partial \overline{(u_ru_\theta)}}{\partial r} 
+ \frac{\partial \overline{(u_zu_\theta)}}{\partial z} 
+ \frac{\overline{u_ru_\theta}}{r}
= 
- \frac{\overline{u_ru_\theta}}{r} - f\overline{u}_r 
+ D_{\nu_m,\theta} \ . 
\end{equation}
Far from the eyewall, the average vertical velocity $\overline{u}_z \approx 0$. 
Thus
\begin{equation} \label{eq:appendix_noSubsidence}
    \overline{u_ru_z} \approx \overline{u_r'u_z'}
     \ ,
    \qquad
    \overline{u_\theta u_z} \approx \overline{u_\theta'u_z'}
     \ .
\end{equation}
Additionally, the following assumption is made
\begin{equation}\label{eq:appendix_nearGroundFollowLogLaw}
    \overline{u_ru_r} \approx \overline{u}_r\overline{u}_r 
     \ ,
    \qquad
    \overline{u_\theta u_r} \approx \overline{u}_\theta\overline{u}_r 
     \ ,
    \qquad
    \overline{u_\theta u_\theta} \approx \overline{u}_\theta\overline{u}_\theta
     \ .
\end{equation}
This assumption might break down close to the surface. 
But close to the surface (and sufficiently far from the eyewall) it is also known that the velocity profiles follow the log-law \citep{powell2003reduced, smith2014existence}. 
Future versions of this procedure that account for variable-eddy viscosity could simply adjust the eddy viscosity so that the velocity follows the log-law near the wall.
The equations become
\begin{equation} \label{eq:appendix_mom_r_subNoSubsidence}
2\overline{u}_r\frac{\partial \overline{u}_r}{\partial r} + 
\frac{\partial \overline{(u_r'u_z')}}{\partial z} + \frac{\overline{u}_r^2}{r}= 
-\left(\frac{V_g^2}{r} + fV_g\right) 
+ \frac{\overline{u}^2_\theta}{r} + f\overline{u_\theta}  
+ D_{\nu_m,r} \ ,
\end{equation}
\begin{equation} \label{eq:appendix_mom_theta_subNoSubsidence}
\overline{u}_\theta\frac{\partial \overline{u}_r}{\partial r} 
+ \overline{u}_r\frac{\partial \overline{u}_\theta}{\partial r} 
+ \frac{\partial \overline{(u_z'u_\theta')}}{\partial z} 
+ 2\frac{\overline{u}_r\overline{u}_\theta}{r}
= - f\overline{u}_r
+ D_{\nu_m,\theta} \ . 
\end{equation}
We now require modelling assumptions for $\partial \overline{u}_\theta/\partial r$ and $\partial \overline{u}_r/\partial r$. 
We follow \cite{bryan2017simple} here.
Assume that, far aloft, the tangential velocity varies radially through a power law
\begin{equation} \label{eq:appendix_GradWindPowerLaw}
    \overline{u}_\theta(r) = V_{max}\left(\frac{r}{r_{max}}\right)^{-n} \ ,
\end{equation}
where $V_{max}$ and $r_{max}$ correspond to a tangential velocity and radial distance within the eyewall. 
Differentiating, we get
\begin{equation} \label{eq:appendix_dvdr_1}
    \frac{d\overline{u}_\theta}{dr}(r) = -\frac{nV_{max}}{r}\left(\frac{r}{r_{max}}\right)^{-n} \ .
\end{equation}
Far aloft, $\overline{u}_\theta(r) = V_g$.
If it is assumed that this radial variation holds throughout the depth of the HBL, then
\begin{equation} \label{eq:appendix_dvdr_2}
    \frac{\partial \overline{u}_\theta}{\partial r}(r,z) \approx -n\frac{V_g}{r} \ .
\end{equation}
To model the term $\partial \overline{u}_r/\partial r$, take the time-average of the continuity equation.
\begin{equation}\label{eq:appendix_continuityTimeAvg}
    \frac{\partial \overline{u}_r}{\partial r} + \frac{\overline{u}_r}{r} + \frac{1}{r}\frac{\partial \overline{u}_\theta}{\partial \theta} + \frac{\partial \overline{u}_z}{\partial z} = 0 \ .
\end{equation}
By statistical homogeneity, $\partial \overline{u}_\theta/\partial \theta = 0$. 
Far from the eyewall, the effect of vertical velocity can also be neglected. Hence
\begin{equation}\label{eq:appendix_dudr}
    \frac{\partial \overline{u}_r}{\partial r} = -\frac{\overline{u}_r}{r} \ .
\end{equation}
Substituting for $\partial\overline{u}_\theta/\partial r$ and $\partial \overline{u}_r/\partial r$, we have
\begin{equation} \label{eq:appendix_mom_r_subModellingAssumps}
\frac{\partial \overline{(u_r'u_z')}}{\partial z} = 
-\left(\frac{V_g^2}{r} + fV_g\right) 
+ \frac{\overline{u}^2_\theta}{r} 
+ \frac{\overline{u}_r^2}{r}
+ f\overline{u}_\theta  
+ D_{\nu_m,r} \ ,
\end{equation}
\begin{equation} \label{eq:appendix_mom_theta_subModellingAssumps}
\frac{\partial \overline{(u_z'u_\theta')}}{\partial z} 
= 
-\frac{\overline{u}_r\overline{u}_\theta}{r}
- f\overline{u}_r 
+ n\overline{u}_r\frac{V_g}{r}
+ D_{\nu_m,\theta} \ . 
\end{equation}
Model the turbulent stresses using a Boussinesq assumption (and absorb the viscous stresses into these). 
At a specific radial distance $r = R$
\begin{eqnarray} \label{eq:appendix_mom_r_subBoussinesq}
\frac{\partial}{\partial z}\left(K\frac{\partial \overline{u}_r}{\partial z} \right) &=& 
\left(\frac{V_g^2}{R} + fV_g\right) 
- \frac{\overline{u}^2_\theta}{R} 
- \frac{\overline{u}_r^2}{R}
- f\overline{u}_\theta \ ,  
\\
\label{eq:appendix_mom_theta_subBoussinesq}
\frac{\partial}{\partial z}\left(K\frac{\partial \overline{u}_\theta}{\partial z} \right)
&=&
\frac{\overline{u}_r\overline{u}_\theta}{R}
+ f\overline{u}_r 
- n\overline{u}_r\frac{V_g}{R} \ .
\end{eqnarray} 
This is a system of two equations has two unknowns $\overline{u}_r$ and $\overline{u}_\theta$, which are functions only of the vertical coordinate $z$. 
The equations to be solved are thus
\begin{eqnarray} \label{appendix_toBeSolvedDimensional_1}
\frac{d}{d z}\left(K\frac{d \overline{u}_r}{d z} \right) &=& 
\left(\frac{V_g^2}{R} + fV_g\right) 
- \frac{\overline{u}^2_\theta}{R} 
- \frac{\overline{u}_r^2}{R}
- f\overline{u}_\theta  
\\
\label{appendix_toBeSolvedDimensional_2}
\frac{d}{dz}\left(K\frac{d \overline{u}_\theta}{d z} \right)
&=& 
\frac{\overline{u}_r\overline{u}_\theta}{R}
+ f\overline{u}_r 
- n\overline{u}_r\frac{V_g}{R}
\end{eqnarray}
\subsection{Appendix 2: Numerical solver}\label{section:appendixNumSoln}
\begin{figure}[htpb]
     \centering
     \begin{subfigure}{0.36\textwidth}
         \centering
    \includegraphics[width=\textwidth]{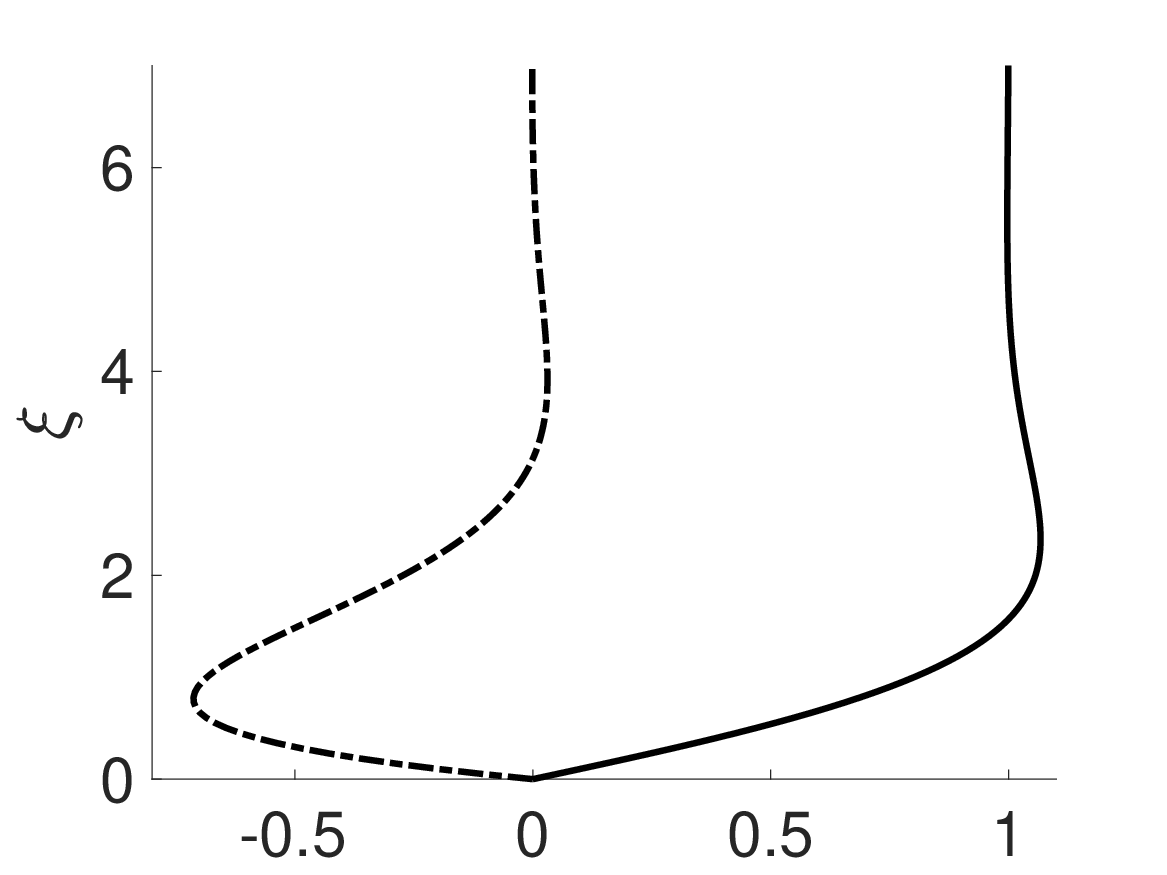}
         \caption*{(a)}
     \end{subfigure}%
     \begin{subfigure}{0.36\textwidth}
         \centering
        \includegraphics[width=\textwidth]{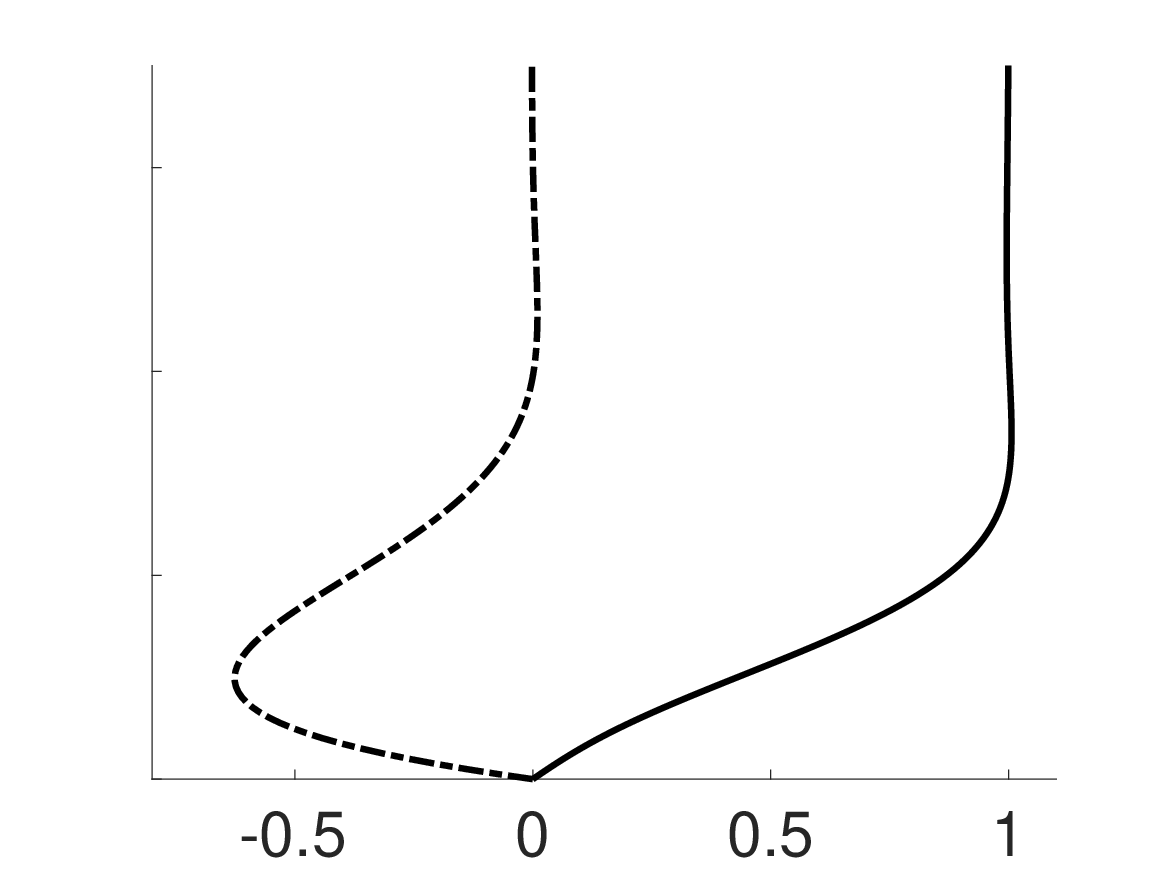}
         \caption*{(b)}
     \end{subfigure}%
     \begin{subfigure}{0.36\textwidth}
         \centering
        \includegraphics[width=\textwidth]{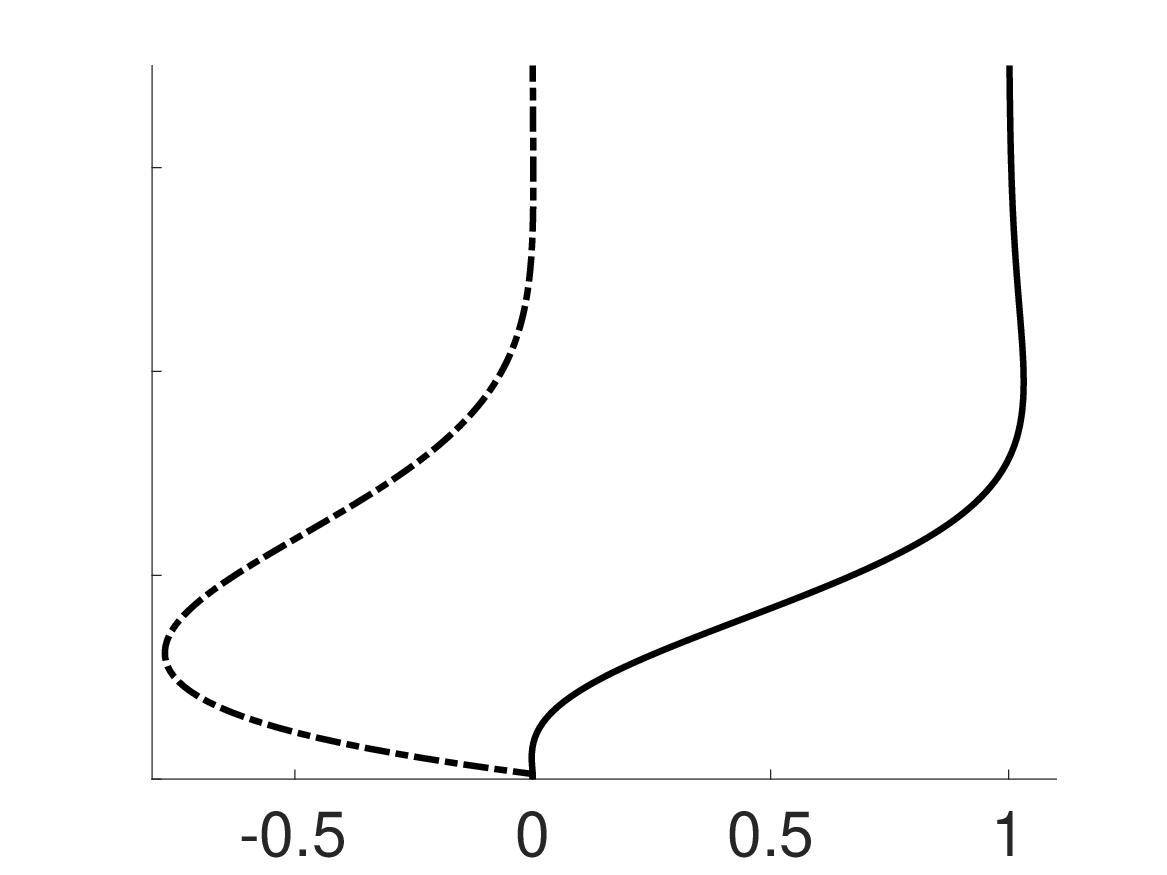}
         \caption*{(c)}
     \end{subfigure}
    \caption{\label{fig:appendix_solutionSwitchMechanism}Snapshots from the pseudo-time evolution of the numerical solution of case 3 of Table 1. Dot-dashed lines are $u$ and solid lines are $1+v$.}
\end{figure}
To solve the system of ordinary differential equations numerically, we use a finite difference based solver. 
A pseudo-time term is added and time marching is performed with a 3\textsuperscript{rd} order Runge-Kutta method. We use a domain size from $\xi = 0$ to $\xi = 10$ and 200 grid points to discretize the domain. 
The initial condition is set to be the linear solution
\begin{eqnarray}\label{eq:appendix_initialCond}
    u_{\text{initial}} &=&  -\frac{\tilde{\alpha}}{2}{\mathrm{e}}^{-\xi}\sin(\xi) \ , \\
    v_{\text{initial}} &=&  -{\mathrm{e}}^{-\xi}\cos(\xi) + \widetilde{G_z}\xi \ .
\end{eqnarray}
First and second derivatives are based on 5\textsuperscript{th} and 6\textsuperscript{th} order WENO schemes respectively.
The WENO based derivatives are capable of capturing sharp gradients and offers greater numerical stability compared to simpler finite differences based derivatives. 
This is especially useful here since there is a significant deviation of the steady state solution for large $n$ from the initial condition, which caused a divergence of the solution when the simpler finite differences based derivatives were used. 

Figure~\ref{fig:appendix_solutionSwitchMechanism} shows snapshots at every 2750 steps of pseudo-time of the solution of case 3 of Table~\ref{tab:n_cases}. 
We notice that, first, the nonlinearities tend to make the gradient of the tangential velocity near the bottom boundary less steep. 
Once the curve becomes approximately perpendicular to the boundary (Fig.~\ref{fig:appendix_solutionSwitchMechanism},c), there is a sharp switch of the tangential velocity curve to the other stable solution (cf.~\eqref{eq:algebraicEqns}). 
This behaviour where the nonlinearities tend to make the gradient of the tangential velocity near the ground smaller is also seen in cases without a switch of solution (Fig.~\ref{fig:n_cases_anlnumSoln}).
This also suggests that the critical value of $n$ could be that for which the slope of the tangential velocity solution is exactly zero at the ground.

The $n$ for which the switch in solution is seen can be postponed if the velocity gradient near the ground is forced to be larger. 
This could be due, for example, to a variable eddy viscosity or a slip boundary condition.
Consider the eddy viscosity profile
\begin{equation} \label{eq:appendix_eddyViscProfile}
    K(z) = 0.4z\left(1-\frac{z}{L}\right)^4 \ .
\end{equation} 
Figure~\ref{fig:appendix_varEddViscShearStress}(a) shows the numerical solution to  \eqref{eq:toBeSolvedDimensional1} and \eqref{eq:toBeSolvedDimensional2} with $n = 0.6$ and using $K(\xi)$ from \eqref{eq:appendix_eddyViscProfile} (with $L=3000$ m) instead of a constant value. The values of the other constants are those specified in Table~\ref{tab:dimValues}.
We see that even for $n=0.6$, the profiles continue to be physical. 
If we push to a larger value of $n$, the switch in solution is seen again, through a similar mechanism as in Fig.~\ref{fig:appendix_solutionSwitchMechanism}. 
By trial and error, the transitional value of $n\approx 0.67$.
\begin{figure}[htpb]
    \centering
    \begin{subfigure}[b]{0.49\textwidth} 
        \centering
        \includegraphics[width=\textwidth]{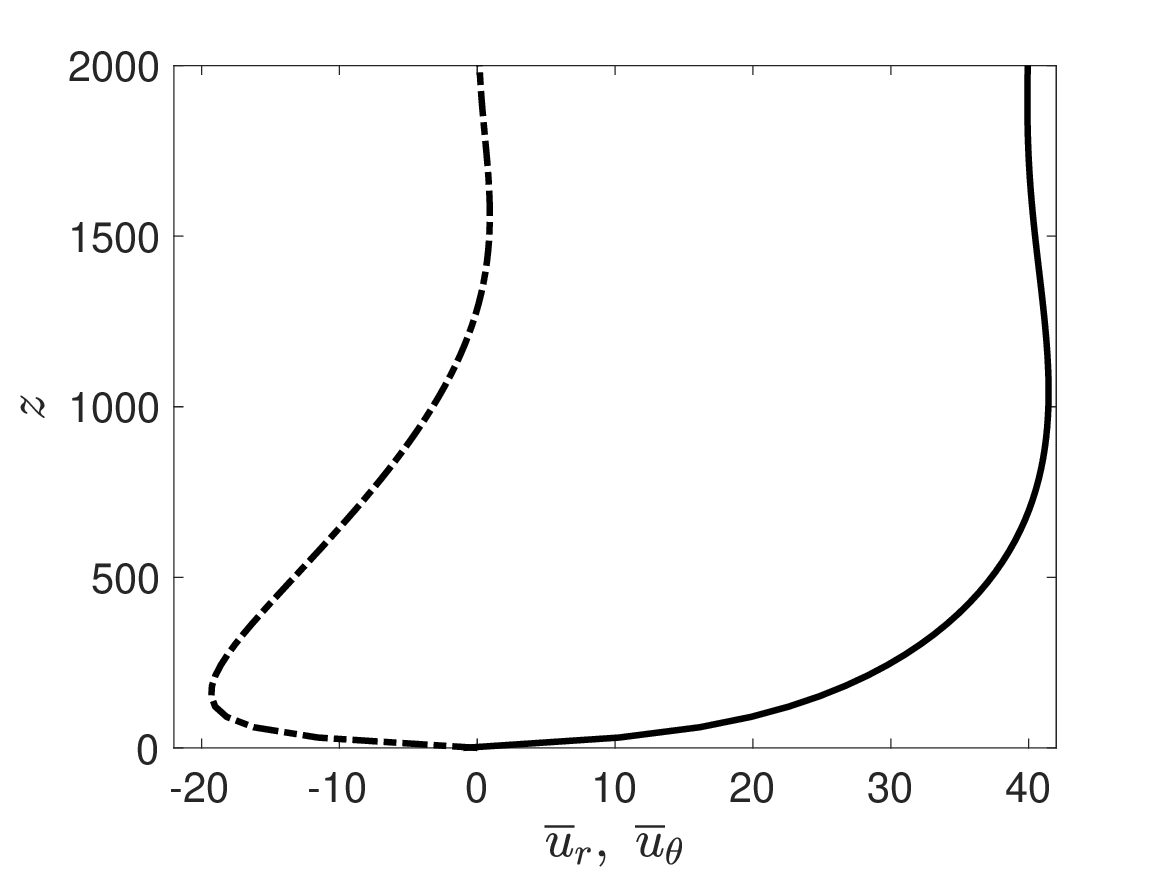}
        \caption*{(a)}  
    \end{subfigure}%
    \hfill  
    \begin{subfigure}[b]{0.49\textwidth}  
        \centering
        \includegraphics[width=\textwidth]{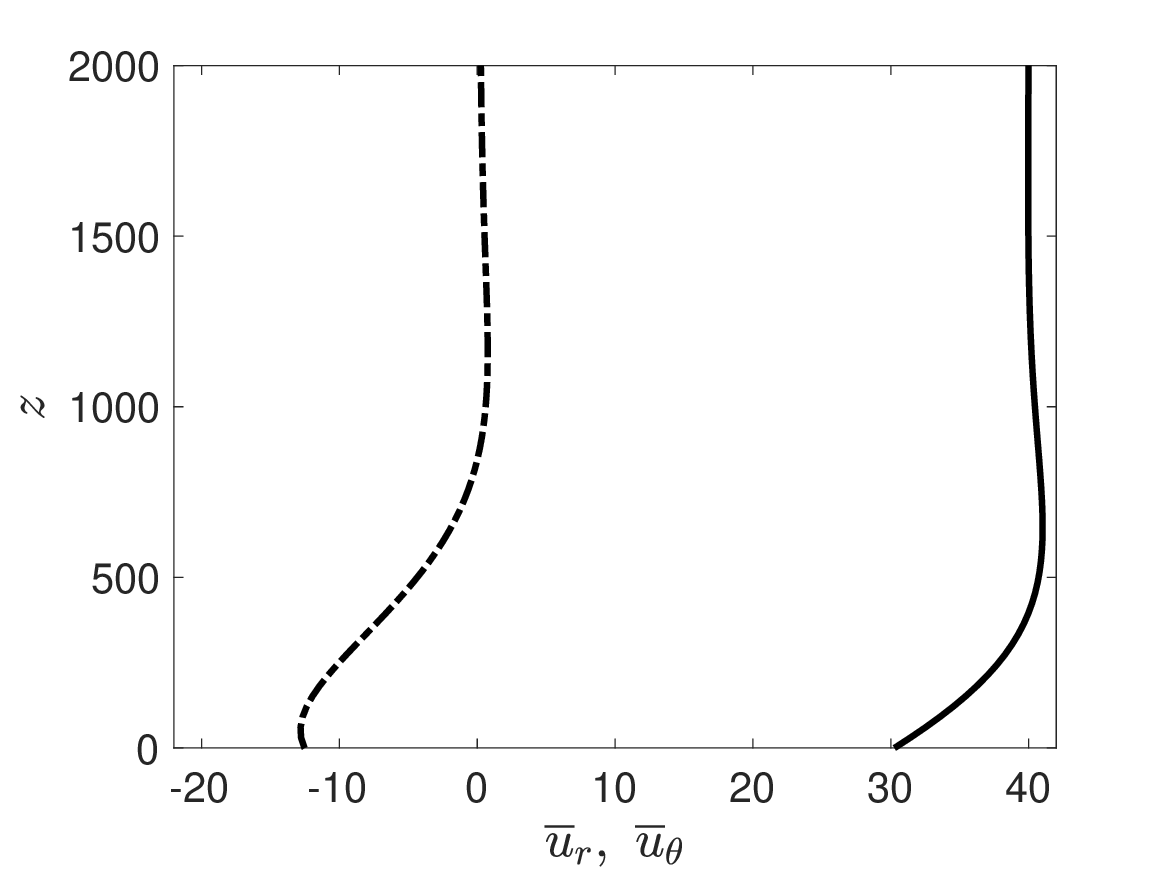}
        \caption*{(b)}  
    \end{subfigure}
    \caption{Numerical solution for $n=0.6$ with (a) variable eddy viscosity (b) a slip boundary condition. Dot-dashed lines are the radial velocity and solid lines are the tangential velocity.}
    \label{fig:appendix_varEddViscShearStress}
\end{figure}
Figure~\ref{fig:appendix_varEddViscShearStress}(b) plots the numerical solution to \eqref{eq:toBeSolvedDimensional1} and \eqref{eq:toBeSolvedDimensional2} again with $n = 0.6$, but using a slip boundary condition at the ground instead of a no-slip condition, as in \citet{kepert2001dynamics}. 
Once again the profiles remain physical.
The switch occurs for $n\gtrapprox 0.69$.
If we choose a larger value of $C=0.02$, the switch is seen at a smaller value of $n \approx 0.55$.
\subsection{Appendix 3: Extension to linearly varying gradient wind} 
\label{section:appendixlinVarGradWind}
Consider a linearly varying gradient wind
\begin{equation}\label{eq:appendix_linVarGradWind}
    V_g = G + z \, G_z
\end{equation}
where $G_z < 0$. 
To incorporate the tendency of $v$ to $V_g$ above the inflow layer, we change the boundary conditions to
\begin{equation}\label{eq:appendix_linVarGradWind_dimBCs}
\begin{gathered}
    \overline{u}_r(z = 0) = 0 \ , \qquad \overline{u}_\theta(z = 0) = 0 \ , \qquad 
    \frac{d\overline{u}_r}{dz}\Biggr|_{z \to \infty} = 0 \ , \qquad \frac{d\overline{u}_\theta}{dz}\Biggr|_{z \to \infty} = G_z \ . 
\end{gathered}
\end{equation}
Substituting ~\eqref{eq:appendix_linVarGradWind} in ~\eqref{eq:toBeSolvedDimensional1} and~\eqref{eq:toBeSolvedDimensional2}, and following the same steps for variable changes and non-dimensionalization, the equations become
\begin{eqnarray}\label{eq:appendix_linVarGradWind_nonDimmedEqns1}
        \frac{\partial^2 u}{\partial \xi^2} &=& -\tilde{\alpha}v - \tilde{\gamma}(u^2+v^2) + (\tilde{\alpha}\widetilde{G_z})\xi + (\tilde{\gamma}\widetilde{G_z}^2)\xi^2 \ , \\
        \label{eq:appendix_linVarGradWind_nonDimmedEqns2}
        \frac{\partial^2 v}{\partial \xi^2} &=& \tilde{\beta}u  + \tilde{\gamma}uv
    - (\tilde{\gamma}\widetilde{G_z}n)u\xi \ ,
\end{eqnarray}
and the boundary conditions become
\begin{equation} \label{eq:appendix_linVarGradWind_nonDimmedBCs}
    u(\xi = 0) = 0 \ , \qquad v(\xi = 0) = -1 \ , \qquad 
    \frac{du}{d\xi}\Biggr|_{\xi \to \infty} = 0 \ , \qquad \frac{dv}{d\xi}\Biggr|_{\xi \to \infty} = \widetilde{G_z} \  
\end{equation}
where $\widetilde{G_z} = {G_zH}/{G}$ is the non-dimensional slope of gradient wind. 
$\widetilde{G_z}$ is now an additional non-dimensional parameter in the governing equations.

Including the artificial small parameter $\delta$ as in \S~\ref{section:analyticalSoln}, the governing equations become
\begin{eqnarray} \label{eq:appendix_linVarGradWind_artificialSmallParam1}
        \frac{\partial^2 u}{\partial \xi^2} & = & -\tilde{\alpha}v + (\tilde{\alpha}\widetilde{G_z})\xi- \delta\left(\tilde{\gamma}(u^2+v^2) -(\tilde{\gamma}\widetilde{G_z}^2)\xi^2 \right) \ ,  \\
        \label{eq:appendix_linVarGradWind_artificialSmallParam2}
        \frac{\partial^2 v}{\partial \xi^2} & = & \tilde{\beta}u  + \delta\left(\tilde{\gamma}uv
    - (\tilde{\gamma}\widetilde{G_z}n)u\xi
    \right) \ .
\end{eqnarray}
Expanding the solution using the asymptotic series \ref{eq:seriesExpansion} results in the following systems of equations at the first few orders. The leading order reads
 \begin{eqnarray}\label{eq:appendix_linVarGradWind_orderEqn_0_a}
 \frac{\partial^2 u_0}{\partial \xi^2}  &=& -\tilde{\alpha}v_0 + (\tilde{\alpha}\widetilde{G_z})\xi \ ,\\
 \label{eq:appendix_linVarGradWind_orderEqn_0_b}
\frac{\partial^2 v_0}{\partial \xi^2}   &=&  
\tilde{\beta}u_0 \ , 
\end{eqnarray}
with boundary conditions $u_0(0) = u_0'(\infty) = 0,v_0(0) = -1, v_0'(\infty) = \widetilde{G_z}$. 
The first order correction is
\begin{eqnarray}\label{eq:appendix_linVarGradWind_orderEqn_1_a}
\frac{\partial^2 u_1}{\partial \xi^2}  &=& 
-\tilde{\alpha}v_1 - \tilde{\gamma}(u_0^2 + v_0^2 - \widetilde{G_z}^2\xi^2) \ , \\
\label{eq:appendix_linVarGradWind_orderEqn_1_b}
\frac{\partial^2 v_1}{\partial \xi^2}   &=&  
\tilde{\beta}u_1 + \tilde{\gamma}(u_0v_0-\widetilde{G_z}n\xi u_0) \ ,
\end{eqnarray}
with boundary conditions $u_1(0) = u_1'(\infty) = v_1(0) = v_1'(\infty) = 0$. 
The second order correction reads
\begin{eqnarray}\label{eq:appendix_linVarGradWind_orderEqn_2_a}
\frac{\partial^2 u_2}{\partial \xi^2}  &=&  {-\tilde{\alpha}v_2} - \tilde{\gamma}(2u_0u_1 + 2v_0v_1) \ ,
\\ 
\label{eq:appendix_linVarGradWind_orderEqn_2_b}
\frac{\partial^2 v_2}{\partial \xi^2}   &=&
\tilde{\beta}u_2 + \tilde{\gamma}(u_0v_1 + u_1v_0-\widetilde{G_z}n\xi u_1) \ ,
\end{eqnarray}
with the same boundary conditions as for the first order problem.

These equations contain additional non-homogeneous terms which are products of polynomial terms $\{\xi,\ \xi^2, ...\}$ and the terms $\{\exp(-\xi)\cos(\xi),\ \exp(-\xi)\sin(\xi), ...\}$. 
This calls for a larger set of basis functions in our ansatz, namely
\begin{equation}\label{eq:appendix_linVarGradWind_ansatz}
    \left\{1,\ \xi,\ \xi^2\right\}\times
    \left\{1,\ e^{-\xi},\ e^{-2\xi}\right\}\times
    \left\{1,\ \cos(\xi),
    \ \sin(\xi),\ \cos(2\xi),
    \ \sin(2\xi)
    \right\}\ .
\end{equation}
Substituting the ansatz and solving for the coefficients readily results in the following leading-order solution 
\begin{eqnarray}\label{eq:appendix_linVarGradWind_orderSoln_0u}
    u_0 &=& -\frac{\tilde{\alpha}}{2}{\mathrm{e}}^{-\xi}\sin(\xi) \ , \\
    \label{eq:appendix_linVarGradWind_orderSoln_0v}
    v_0 &=& -{\mathrm{e}}^{-\xi}\cos(\xi) + \widetilde{G_z}\xi \ ,
\end{eqnarray}
and first-order correction  
\begin{equation}\label{eq:appendix_linVarGradWind_orderSoln_1u}
\begin{gathered}    
    u_1(\xi) = 
    \frac{\tilde{\gamma}}{10}\left( 
    \frac{\tilde{\alpha}^2}{4} + 1
    \right)e^{-\xi}\cos(\xi) + 
    \frac{\tilde{\gamma}}{30}e^{-\xi}\sin(\xi)  \\
-
\frac{\tilde{\gamma}}{10}\left( 
    \frac{\tilde{\alpha}^2}{4} + 1
       \right)e^{-2\xi} + 
       \frac{\tilde{\gamma}}{30}\left( 
    -\frac{3}{8}\tilde{\alpha}^2 +2
       \right)e^{-2\xi}\sin(2\xi) + 
\frac{\widetilde{G_z}\tilde{\gamma}\tilde{\alpha}^2}{8}\left(1-n\right)\xi e^{-\xi}\sin(\xi) \ ,
       \end{gathered}
       \end{equation}
       \begin{equation}
       \label{eq:appendix_linVarGradWind_orderSoln_1v}
       \begin{gathered}   
       v_1(\xi) = 
    \frac{\tilde{\gamma}}{15\tilde{\alpha}}e^{-\xi}\cos(\xi) - 
    \frac{\tilde{\gamma}}{5\tilde{\alpha}}\left(\frac{\tilde{\alpha}^2}{4} + 1\right)e^{-\xi}\sin(\xi)\\
- \frac{\tilde{\gamma}}{10\tilde{\alpha}}\left(\frac{\tilde{\alpha}^2}{4} + 1\right)e^{-2\xi} +
\frac{\tilde{\gamma}}{30\tilde{\alpha}}\left(\frac{3\tilde{\alpha}^2}{4} + 1\right)e^{-2\xi}\cos(2\xi) \ .
\end{gathered}
\end{equation} 
Higher order solutions can similarly be obtained. 
We observe that there is no term involving $\widetilde{G_z}$ in $v_1$. 
Even the term involving $\widetilde{G_z}$ in $u_1$ is quite small compared to the other terms and almost no difference is seen in a plot where the term is ignored (not shown). 
Hence one can sufficiently incorporate the effect of $\widetilde{G_z}$ with just the term $\widetilde{G_z}\xi$ in $v_0$ and safely ignore its presence in the corrections.

 \begin{figure}[htpb]
     \centering
     \begin{subfigure}{0.49\textwidth}
         \centering
    \includegraphics[width=\textwidth]{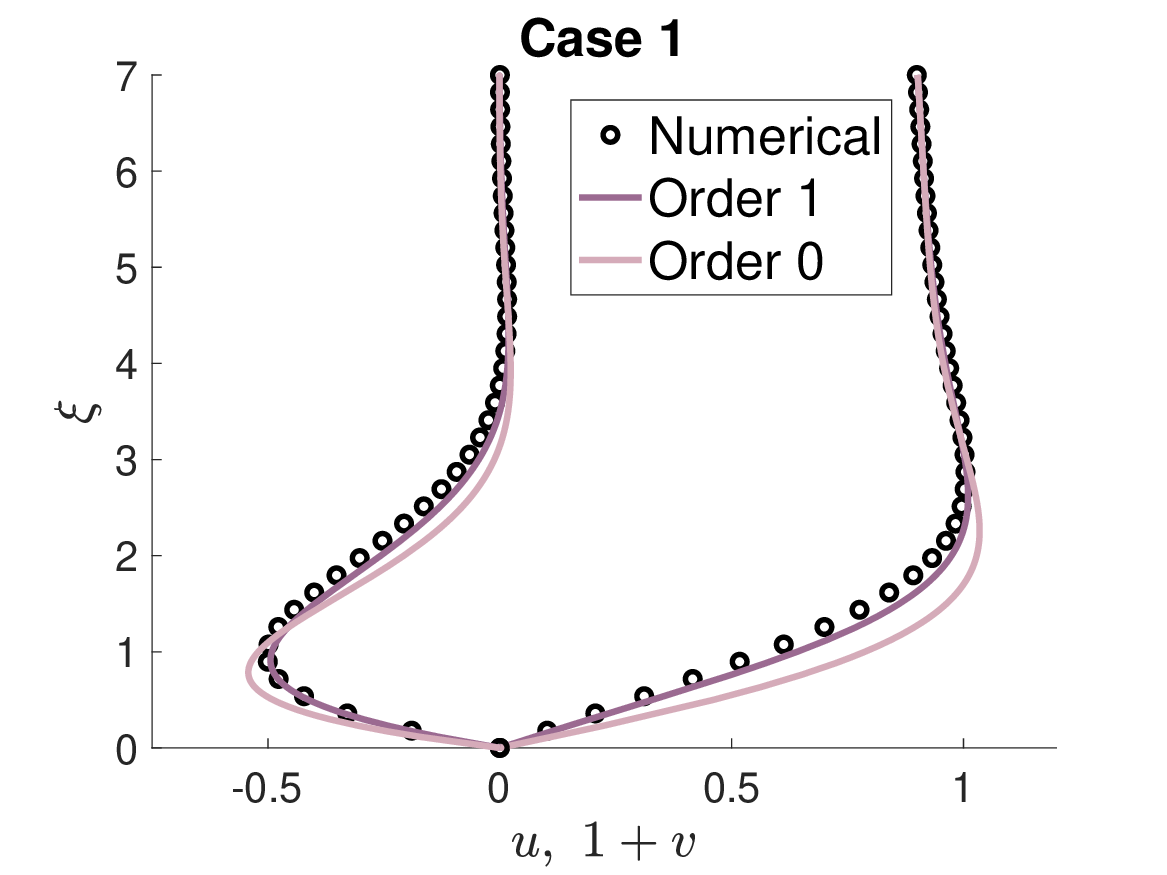}
    \caption*{(a)}
     \end{subfigure}
     \begin{subfigure}{0.49\textwidth}
         \centering
        \includegraphics[width=\textwidth]{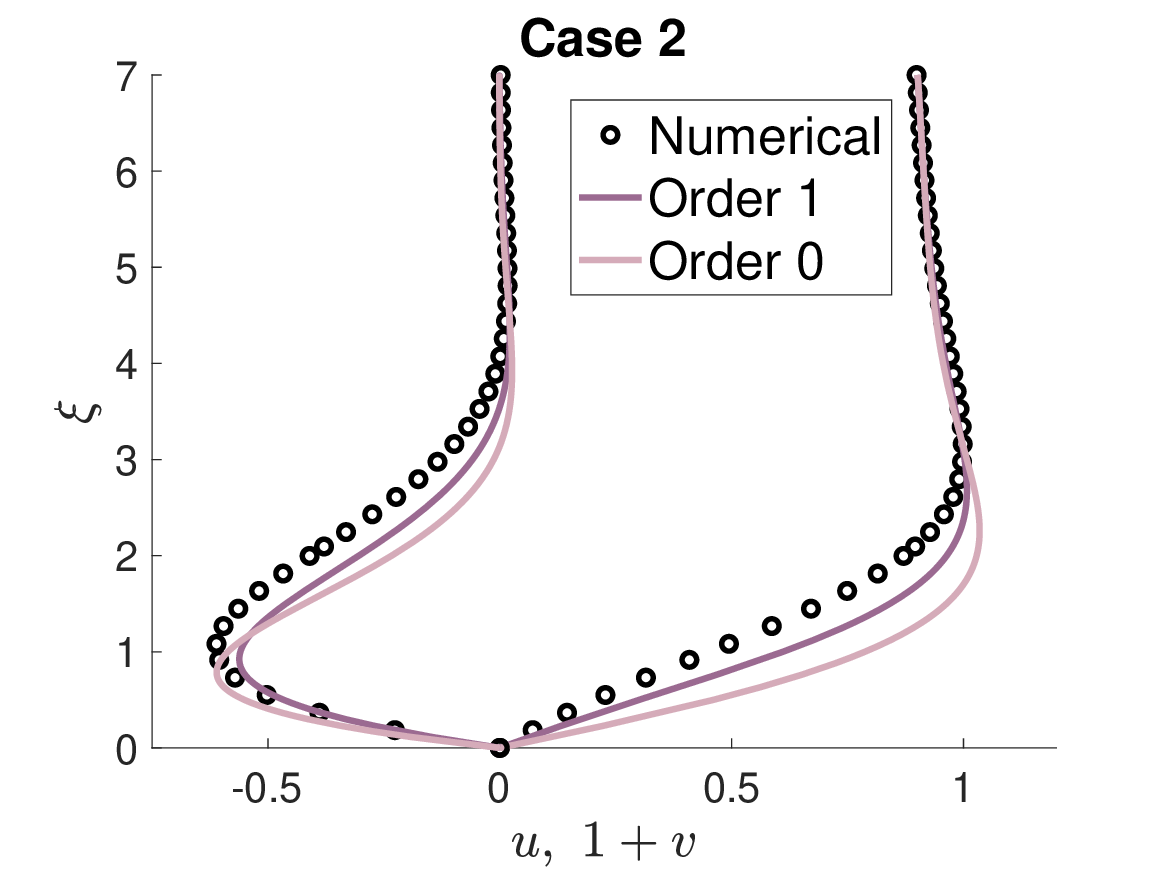}
        \caption*{(b)}
     \end{subfigure}
     \caption{\label{fig:Table2CasedVgdz}Cases 1 and 2 from Table~\ref{tab:n_cases}, with $\widetilde{G_z} = 0.014$.}
\end{figure}

Due to limited observational data, there are very few numerical values reported in the literature for $G_z$. 
Here, we choose a representative value to be that used in \citet{bryan2017simple}, i.e., a decrease of the gradient wind speed of $40$ m s\textsuperscript{-1} over $18$ km. 
Assuming this $G_z$ is for $R = 50$ km and $K = 50$ m\textsuperscript{2} s\textsuperscript{-1} gives $H = 250$ m. Hence $\widetilde{G_z} = 250/18000 = 0.014$.

Figure~\ref{fig:Table2CasedVgdz} shows the comparison between the numerical and analytical solutions for Cases 2 and 4 in Table 2, and with $\widetilde{G_z} = 0.014$. 
We observe, as before, that the solution converges well for relatively smaller $n$, whereas the convergence is slower as $n$ approaches the limiting value $n \approx 0.5$. 
The solution again diverges for larger $n$, with the limit $n\approx 0.5$ not appearing to be affected by the specific value of $\widetilde{G_z}$ (not shown).
\subsection{Appendix 4: Higher order approximation solution forms}
\label{section:appendixhigherOrderSolns}
The second order corrections are

\begin{equation}\label{eq:appendix_higherOrderSoln_u}
    \begin{aligned}
    u_2(\xi) = \tilde{\gamma}^2\Bigg(
 \frac{{\mathrm{e}}^{-\xi} \cos(\xi)}{25600\, \tilde{\alpha}} \left(25\, \tilde{\alpha}^4 + 98\, \tilde{\alpha}^2 + 688\right)
- \frac{{\mathrm{e}}^{-\xi} \sin(\xi)}{115200\, \tilde{\alpha}} \left(135\, \tilde{\alpha}^4 + 210\, \tilde{\alpha}^2 + 368\right)\\
+ \frac{{\mathrm{e}}^{-2 \xi}}{300\, \tilde{\alpha}} \left(\tilde{\alpha}^2 + 4\right)
+ \frac{{\mathrm{e}}^{-2 \xi} \cos(2 \xi)}{2400\, \tilde{\alpha}} \left(3\, \tilde{\alpha}^4 - 4\, \tilde{\alpha}^4 - 64\right) 
+ \frac{{\mathrm{e}}^{-2 \xi} \sin(2 \xi)}{1800\, \tilde{\alpha}} \left(3\, \tilde{\alpha}^2 - 16\right)\\
- \frac{{\mathrm{e}}^{-3 \xi} \cos(\xi)}{76800\, \tilde{\alpha}} \left(171\, \tilde{\alpha}^4 + 422\, \tilde{\alpha}^2 + 1040\right) 
- \frac{{\mathrm{e}}^{-3 \xi} \sin(\xi)}{25600\, \tilde{\alpha}} \left(39\, \tilde{\alpha}^4 + 198\, \tilde{\alpha}^2 + 240\right) \\
- \frac{{\mathrm{e}}^{-3 \xi} \sin(3 \xi)}{76800\, \tilde{\alpha}} \left(27\, \tilde{\alpha}^4 - 26\, \tilde{\alpha}^2 + 144\right)
\Bigg) \ ,
    \end{aligned}
\end{equation}

\begin{equation}\label{eq:appendix_higherOrderSoln_v}
    \begin{aligned}
    v_2(\xi) = 
    \tilde{\gamma}^2\Bigg(
 \frac{{\mathrm{e}}^{-\xi} \cos(\xi)}{115200\, \tilde{\alpha}^2} \left(270\, \tilde{\alpha}^4 + 420\, \tilde{\alpha}^2 + 736\right)
+ \frac{{\mathrm{e}}^{-\xi} \sin(\xi)}{25600\, \tilde{\alpha}^2} \left(50\, \tilde{\alpha}^4 + 196\, \tilde{\alpha}^2 + 1376\right)\\
+ \frac{{\mathrm{e}}^{-2 \xi}}{300\, \tilde{\alpha}^2} \left(\tilde{\alpha}^2 + 4\right)
- \frac{{\mathrm{e}}^{-2 \xi} \cos(2 \xi)}{900\, \tilde{\alpha}^2} \left(3\, \tilde{\alpha}^2 + 4\right) 
+ \frac{{\mathrm{e}}^{-2 \xi} \sin(2 \xi)}{1200\, \tilde{\alpha}^2} \left(3\, \tilde{\alpha}^4 + 16\, \tilde{\alpha}^2 + 16\right)\\
- \frac{{\mathrm{e}}^{-3 \xi} \cos(\xi)}{76800\, \tilde{\alpha}^2} \left(186\, \tilde{\alpha}^4 + 452\, \tilde{\alpha}^2 - 160\right) 
- \frac{{\mathrm{e}}^{-3 \xi} \sin(\xi)}{25600\, \tilde{\alpha}^2} \left(14\, \tilde{\alpha}^4 - 132\, \tilde{\alpha}^2 + 160\right) \\
- \frac{{\mathrm{e}}^{-3 \xi} \cos(3 \xi)}{76800\, \tilde{\alpha}^2} \left(6\, \tilde{\alpha}^4 + 172\, \tilde{\alpha}^2 + 32\right)
\Bigg) \ .
    \end{aligned}
\end{equation}
Higher order corrections are similarly obtained, but involve many more cross terms.
\end{document}